\documentclass[aps,
prfluids,
final
singlecolumn,
singlespace,
groupedaddress,
notitlepage 
]{revtex4-1}

\usepackage{amsthm}
\usepackage{subfigure}
\usepackage{graphicx}
\usepackage{epsfig}
\usepackage{epstopdf}
\usepackage{color}
\usepackage{natbib}
\usepackage{amsmath}
\usepackage{amssymb}
\usepackage{hyperref}

\begin{document}

\title{Confinement induced control of similarity solutions in premelting dynamics and other thin film problems}


\author{Satyajit Pramanik}
\email[]{satyajit.math16@gmail.com}
\affiliation{Nordita, Royal Institute of Technology and Stockholm University, Stockholm, Sweden}

\author{John S. Wettlaufer}
\email[]{john.wettlaufer@yale.edu}
\affiliation{Yale University, New Haven, Connecticut, USA}
\affiliation{Mathematical Institute, University of Oxford, Oxford, UK}
\affiliation{Nordita, Royal Institute of Technology and Stockholm University, Stockholm, Sweden}


\date{\today}

\begin{abstract}
We study the combined effects of nonlocal elasticity and confinement induced ordering on the dynamics of thermomolecular pressure gradient driven premelted films bound by an elastic membrane. The confinement induced ordering is modeled using a film thickness dependent viscosity. When there is no confinement induced ordering, we recover the similarity solution for the evolution of the elastic membrane, which exhibits an infinite sequence of oscillations. However, when the confinement induced viscosity is comparable to the bulk viscosity, the numerical solutions of the full system reveal the conditions under which the oscillations and similarity solutions vanish. Implications of our results for general thermomechanical dynamics, frost heave observations and cryogenic cell preservation are discussed. 
\textcolor{black}{Finally, through its influence on the viscosity, the confinement effect implicitly introduces a new universal length scale into the volume flux.  Thus, there are a host of thin film problems, from droplet breakup to wetting/dewetting dynamics, whose properties (similarity solutions, regularization, and compact support) will change under the action of the confinement effect. Therefore, our study suggests revisiting the mathematical structure and experimental implications of a wide range of problems within the framework of the confinement effect.}
\end{abstract}


\maketitle


\section{Introduction\label{sec:intro}} 
The free surfaces of a wide range of materials (e.g., rare gas solids, metals, quantum and molecular solids) are observed to have a thermodynamically stable thin liquid film that disjoins the solid from its vapor phase, another grain of solid, or another material entirely. This extension of the equilibrium domain of the melt phase into the solid region of the bulk phase diagram is known as \emph{interfacial premelting}, the detailed manifestation of which depends on the dominant polarization and surface forces in the system \cite{Wilenetal1995, Dash2006, Wettlaufer2006}. The same general phenomena, whereby volume-volume interactions that are extensive with the system volume compete with those that are extensive with interfacial area, are at play in, amongst many other settings, wetting, biomaterials, ceramics, colloids, and tribology \cite{SchickLesHouches, Safranbook, French:2010, Jacobbook}. Here we focus on \emph{premelting dynamics}, in a setting in which the melt phase appears between the solid and a flexible wall and flows from high to low temperatures under the influence of thermomolecular pressure gradients \cite{Wilen1995, Dash2006, Wettlaufer2006}. These thin film flows have been studied theoretically and experimentally on ice \cite{Dash2006, Wettlaufer2006, Wilen1995, Wettlaufer1995, Wettlaufer1996, Rempel2004}, argon \cite{Zhu2000}, benzene \cite{Taber1929} and helium \cite{Mizusaki1995}. Importantly, because the liquid phase is adjacent to its own solid, the region near the latter is endowed with solid-like ordering over a length scale that is system specific. It is this ``confinement-induced'' ordering that we examine in the context of premelting dynamics, extending a recent study in which the membrane imposed a local elastic hoop stress \cite{Pramanik2017}. 

A host of effects are responsible for confined liquids exhibiting properties that differ from their bulk counterparts \cite{SchickLesHouches, Evans1990, Hugo2001}. For example, molecular dynamics simulations of confined nanometric liquids show wall normal fluid layering as a function of temperature, the nature of the wall-fluid and the fluid-fluid interactions, and, for temperatures close to the melting point, transitions between shear thinning and shear thickening are observed \cite{Koplik1995}. Interfacially premelted water on ice surfaces exhibits mobility attenuation several orders of magnitude larger than that of bulk water \cite{Pittenger2001}, which is of relevance, for example to the kinetic friction of ice surfaces \cite{Butt1998, Dash2003}. Recent experiments on the dynamics of premelted water on ice surfaces determined the ratio of surface tension-to-shear viscosity, from which it was inferred that the shear viscosity of premelted water is ten times that of bulk supercooled water \cite{Murata2015}. This observation is a central motivation of our work. 

Experiments show that the shear viscosity of thin liquid films of different materials are considerably greater than their bulk values \cite{Granick1991, Dhinojwala1997, Raviv2002, Zhu2003, Zhu2004, Major2006, Bureau2010}. For example, the effective viscosity of nanometric thin water films confined between different surfaces increases from the bulk value by almost seven orders of magnitude \cite{Dhinojwala1997, Major2006}. The increase in effective viscosity of molecularly thin films of octamethylcyclotetrasiloxane  as the films thinned was shown to depend on film thickness as a power law \cite{Bureau2010}. Confinement-induced enhancement of the coefficient of shear viscosity has also been reported for dodecane \cite{Granick1991} and alkanes \cite{Zhu2004}, and depends upon the quenching rate and the degree of perfection of the mica surfaces. Contrary to these systems, aqueous salt solutions can evidently retain their bulk properties at nanometer scales \cite{Raviv2002}. 

Despite the host of implications of the confinement effects of liquids on the dynamical consequences of premelted liquid films, there has been a single study along these lines. In a recent paper, motivated by the experiments discussed above, we treated the premelted liquid viscosity as a power law function of the film thickness \cite{Pramanik2017}. We studied the influence of confinement effects in premelting dynamics on ice surfaces in elastic capillary tubes with a local hoop stress coupling, a setting that had previously been examined without confinement effects \cite{Wettlaufer1995}. 

Our purpose here is to explore the coupled effects of nonlocal elastic deformation of a membrane confining a thin premelted film on the surface of ice, and confinement enhanced viscosity, in a geometry studied previously in the absence of confinement effects \cite{Wettlaufer1996}. In the absence of confinement effects one finds a family of fourth order parabolic partial differential equations (PDEs) that depend on the nature of the interactions responsible for the existence of the premelted film, and there is an associated family of ordinary differential equations (ODEs) with a family of similarity solutions \cite{Wettlaufer1996}. Here, amongst other things, we find that there exists a finite range of the power-law exponents of the confinement induced viscosity model that impose a length scale on the problem and thereby lead to the breakdown of the similarity solutions.  \textcolor{black}{This has general consequences for a wide range of thin film problems.}

Of particular relevance are the consequences for continued flow of liquid at low temperatures where confinement effects should be at play and where no quantitative experiments have been undertaken. Settings where this may play out include the flow of unfrozen water in terrestrial, planetary and engineering environments \cite{Kessler2003, Andersland2004, Rempel2007, Peterson2008, Tamppari:2012}, and during biological freezing \cite{Rubinsky2003, Hawthorn2016}. For example, cardiological treatment is often facilitated by lowering temperatures to lower metabolic rates and during \emph{cryogenic preservation} cells are preserved below freezing point of water \cite{Rubinsky2003}. Thus, mechanical damage to cells during freeze-thaw cycling is a fundamental challenge \cite{Ishiguro1994}, for which many approaches are taken. For example, optimal cryogenic cell preservation may be facilitated by introduction of \emph{cryoprotectants}, such as ice-binding proteins \cite{Bar-Dolev:2012}, and controlled rates of solidification \cite{Rubinsky1988}, all of which will be influenced by the slow flow of confined premelted water abutting compliant walls. 

Finally, we note that similar higher-order diffusion equations govern the dynamics of viscous currents down an incline \cite{Huppert1982, Troian1989}, and a spate of other thin-film settings \cite{Oron:1997, Craster:2009} including models of Hele-Shaw flows \cite{Constantin:1993}, viscous gravity currents between a rigid surface and a deformable elastic sheet \cite{Hewitt2015} and many others too numerous to allow for an exhaustive list. Motivated by these problems, there is a literature on the existence, regularity, and the finite time blow up of nonlinear, degenerate, fourth-order parabolic differential equations (e.g., \cite{Bertozzi2000} and references therein). Mathematically our study falls under the general rubric of this class of equations, but physically it is a unique thin film system, being driven as it is by thermomolecular pressure gradients and phase changes.  \textcolor{black}{Importantly, because of the new length scale introduced by the confinement effect, its inclusion will influence {\em all} of these problems in a fundamental manner.} 

In Section \ref{sec:curvature}, we briefly review the model and the non-dimensional formulation. In Section \ref{sec:similarity_sol}, we discuss the similarity solutions in the two limiting cases of viscosity dominiated by either the bulk or the confinement induced values. Section \ref{sec:Results} examines the membrane deformation for different parameter values, and the numerical results are compared with experiments in Section \ref{sec:summary}, before concluding. 

\begin{figure}[hbtp]
\centering
\includegraphics[scale=0.4]{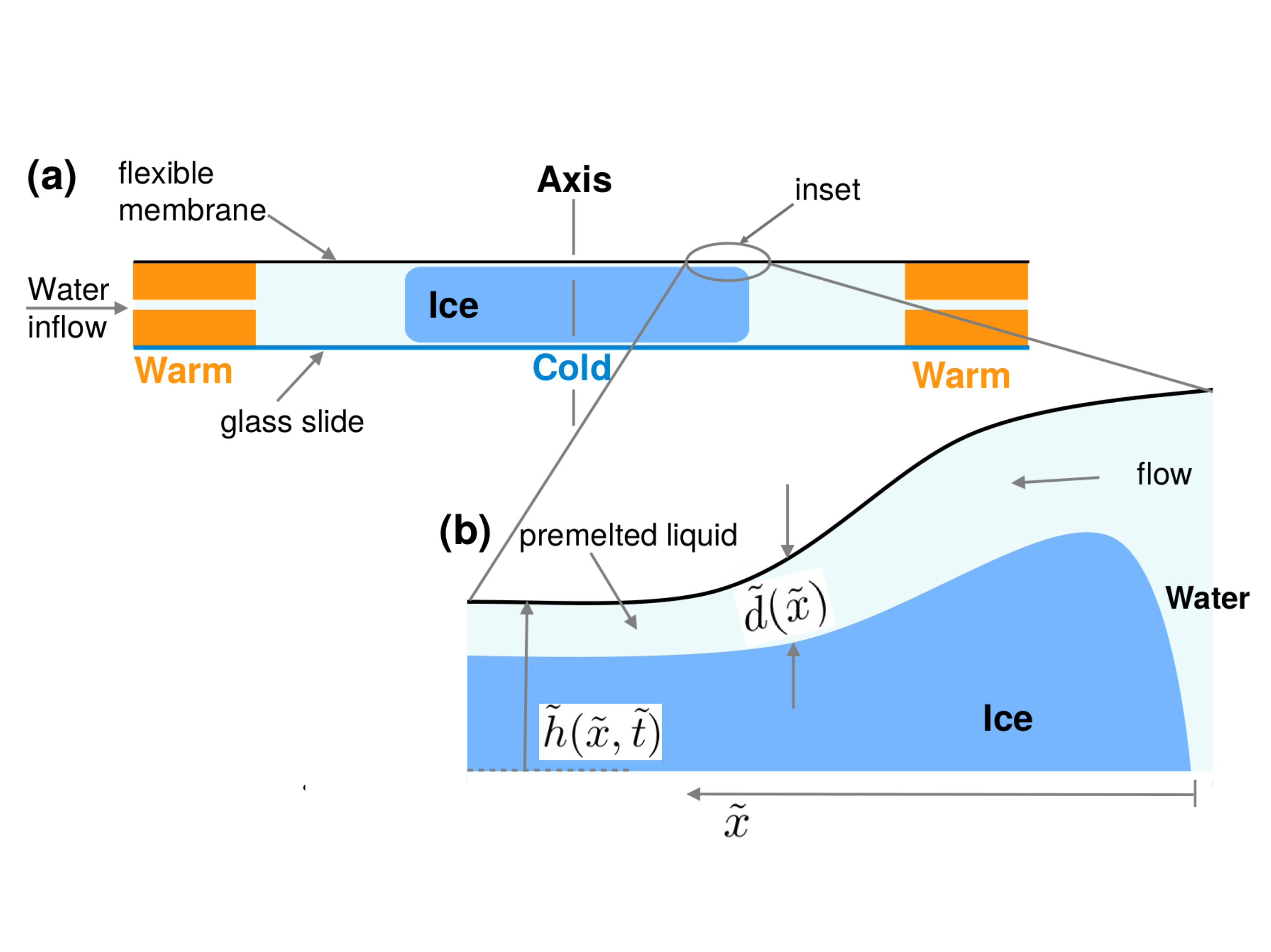}
\caption{(a) Schematic of the experimental set up of Wilen and Dash \cite{Wilen1995}, taken from \cite{Wettlaufer2006}. (b) Inset depicting thermomolecular pressure gradient driven dynamics of a premelted film.} 
\label{fig:schematic}
\end{figure} 

\section{Premelting dynamics and thermodynamics\label{sec:curvature}} 
In Figure \ref{fig:schematic} we show a schematic of the experimental setting that motivates our study. The interface between a single ice crystal and a flexible polymer membrane is subject to a temperature gradient parallel to the interface \cite{Wilen1995}. The membrane/ice interface is disjoined by a premelted layer of water, which flows from high to low temperature. Because at each temperature the premelted liquid film has a thermodynamically required thickness, determined by Eq. \eqref{eq:film_thickness} below, mass conservation insures that a gradient in the liquid volume flux will lead to solidification and hence the deformation of the height of the membrane, $\tilde{h}(\tilde{x}, \tilde{t})$, measured relative to its initial reference height. Our choice of coordinates places the bulk solid-liquid interface at $\tilde{x} = 0$, where $\tilde{x}$ increases with decreasing temperature. Because the deformation of the membrane height is much smaller than the disk radius, a one dimensional model is appropriate and was first developed to analyze experiments using the bulk liquid viscosity \cite{Wettlaufer1996}. We revisit this model, but with a viscosity that depends on the thickness of the premelted film. 

The free energy of the ice/film/membrane system is given by the grand potential 
\begin{equation}
\label{eq:potential} 
\Omega = -P_l V_l - P_s V_s + \mathcal{I} (\tilde{d}), 
\end{equation}
where $ \mathcal{I}(\tilde{d}) = \left[ \Delta \sigma f(\tilde{d}) + \sigma_{sw} \right] A_i $ is the effective interfacial free energy, 
$\tilde{d}$ is the film thickness, 
$ A_i $ is the interfacial area, and $P$ and $V$ are the pressure and volume.  The difference between the dry and wet interfacial free energies is $ \Delta \sigma = \sigma_{sl} + \sigma_{lw} - \sigma_{sw} $,  where the $ \sigma $'s are the solid-liquid ($sl$), liquid-wall ($lw$), and solid-wall ($lw$) values, where $\sigma_{sw} > \sigma_{sl} + \sigma_{lw}$ is a sufficient condition for the interface to premelt \cite{Wettlaufer1996}. When the polarization forces underlying the wetting behavior are power law interactions, then $f(\tilde{d})$ is written as 
\begin{equation}
\label{eq:powerlaw_interaction} 
f(\tilde{d}) = 1 - \left( \frac{ d_0 }{ \tilde{d} } \right)^{n-1}, 
\end{equation}
where $n$ depends on the nature of interactions attracting the liquid to the solid and $d_0$ in the order of molecular diameter. For example, non-retarded (retarded) van der Waals interactions are represented by $n = 3$ ($n = 4$). We minimize $\Omega$ at fixed temperature $T$, close to the bulk melting temperature $T_m$, total volume and chemical potential to obtain two key results. Firstly \cite{Wettlaufer1996},  
\begin{equation}
\label{eq:film_thickness} 
\tilde{d} = \left[ - \frac{ (n-1) d_0^{n-1} \Delta \sigma }{ \rho_l q_m } \right]^{1/n} t_r^{-1/n} \equiv \lambda_n t_r^{-1/n}, 
\end{equation}
where, $t_r = (T_m - T)/T_m$ is the reduced temperature, $\rho_l$ is the liquid density, and $q_m$ the latent heat of fusion. Secondly, the pressure difference between the liquid film and the bulk solid, 
\begin{equation}
\label{eq:disjoining_pressure}
P_l - P_s = \Delta \sigma (n - 1) d_0^{n-1} \tilde{d}^{-n}, 
\end{equation}
which reflects the polarization forces attracting the liquid to the solid. Combining the temperature dependent film thickness \eqref{eq:film_thickness} and the disjoining pressure \eqref{eq:disjoining_pressure} we obtain the melt pressure as a function of the temperature 
\begin{equation}
\label{eq:thermomlecular_pressure}
P_l = P_s - \rho_l q_m t_r, 
\end{equation}
independent of the nature of interactions, characterized by $n$. Clearly, when fixing the solid pressure, $P_s$, and imposing a temperature gradient parallel to the premelted surface, a pressure gradient drives thin film flow towards lower temperatures \cite{Wettlaufer1995, Wettlaufer1996}, as shown in Figure \ref{fig:schematic}.

As the temperature decreases equation \eqref{eq:film_thickness} shows that the film thins. However, as the temperature becomes very low, this does not capture the fact that depending upon crystallographic orientation of the solid the film properties are intermediate between a liquid and a solid (for the ice-water interface, see \cite{Nada:1997, Nada:2000, Nada:2016}). Therefore, the ordering effect of the solid on the liquid may have the dynamical consequences as the premelted film continually thins at low temperatures. This proximity effect on the liquid film viscosity is modeled as  
\begin{equation}
\label{eq:powerlaw_visco_d}
\tilde{ \eta }( \tilde{ d } ) = \eta_0 - \left( \eta_0 - \eta_b \right)\left[ 1 - \left( \frac{ d_0 }{ \tilde{ d } } \right)^{\gamma} \right], 
\end{equation}
where $\gamma > 0$ such that $\tilde{ \eta }( \tilde{ d } ) \to \eta_b$ (bulk viscosity) as $\tilde{ d } \to \infty$, and $\tilde{ \eta }( \tilde{ d } ) \to \eta_0$ when $\tilde{ d } \to d_0^{+}$, and the continuum approximation breaks down \cite{Pramanik2017}. For $\tilde{ d } = \lambda_n t_r^{-1/n}$ and a constant temperature gradient of $\mathcal{G}$, the reduced temperature depends on position through
\begin{equation}
\label{eq:reduced_temperature}
t_r = \frac{ \mathcal{G}\tilde{ x } }{T_m}.  
\end{equation}
Thus, the viscosity in \eqref{eq:powerlaw_visco_d} can be expressed as a function of the position as
\begin{equation}
\label{eq:powerlaw_visco_x}
\tilde{ \eta }( \tilde{ x } ) = \eta_0 - \left( \eta_0 - \eta_b \right) \left[ 1 - \left\{ \frac{ d_0 }{ \lambda_n } \left( \frac{ \mathcal{G} }{ T_m } \right)^{1/n} \right\}^{\gamma} \tilde{ x }^{\gamma/n} \right], 
\end{equation}
and, as we shall see, this position dependence creates an internal length scale in the problem that leads to a parameter dependent breakdown in the family of similarity solutions that are found when the viscosity is constant \cite{Wettlaufer1996}. 

In the lubrication approximation the volume flux $Q$, per unit breadth through the film of thickness $\tilde{d}$ given by \eqref{eq:film_thickness}, is
\begin{equation}
\label{eq:Q}
Q(\tilde{x}) = - \frac{ \tilde{d}^3 }{ 12 \tilde{\eta}(\tilde{x}) } \partial_{ \tilde{x} } P_l.  
\end{equation}
The pressure $P_s$ exerted on the solid phase by the membrane is proportional to the membrane curvature $\kappa$, which in the small slope approximation is $\kappa \approx \tilde{h}_{ \tilde{x} \tilde{x} }$, giving 
\begin{equation}
\label{eq:Young-Laplace}
P_s = -\tilde{\sigma} \tilde{h}_{ \tilde{x} \tilde{x}}, 
\end{equation}
where the membrane surface tension is $\tilde{\sigma}$. Combining \eqref{eq:disjoining_pressure}, \eqref{eq:thermomlecular_pressure}, and \eqref{eq:reduced_temperature} -- \eqref{eq:Young-Laplace}, mass conservation is written as
\begin{equation}
\label{eq:mass_conservation} 
\partial_{ \tilde{ t } } \tilde{ h } + \Gamma_n \partial_{ \tilde{ x } } \left[ \frac{ \tilde{x}^{-3/n} }{ \tilde{\eta}(\tilde{ x })/\eta_b } \left( 1 + \tilde{ \alpha } \tilde{ h }_{ \tilde{x} \tilde{x} \tilde{x} } \right) \right] = 0, 
\end{equation}
where 
\begin{equation}
\label{eq:Gamma_alpha}
\Gamma_n = \frac{\lambda_n^3 \rho_l q_m }{12 \eta_b}\left(\frac{T_m}{\mathcal{G}}\right)^{(3/n-1)} \quad\textrm{and}\quad \tilde{\alpha} = \frac{\widetilde{\sigma}T_m}{\rho_s q_m \mathcal{G}},  
\end{equation}
Because the temperature gradient $\mathcal{G}$ is a constant, at each position $\tilde{ x }$ thermodynamic equilibrium requires the film thickness to satisfy \eqref{eq:film_thickness}. Therefore, when the thermomolecular pressure driven flow brings liquid in excess of the equilibrium value, the evolution equation \eqref{eq:mass_conservation} describes how the membrane must deform in response to the solidification of the premelted fluid, as seen schematically in Figure \ref{fig:schematic}.

The quantitative properties of the premelting dynamics with the proximity effect of \eqref{eq:powerlaw_visco_d} differ from the bulk viscosity case, but not in the high temperature regime that has thus far been experimentally accessible \cite{Wilen1995}. Hence, our calculations here are intended to motivate experiments at low temperatures and long time scales where the proximity effect becomes important. 

\subsection{Non-dimensionalization\label{subsec:nondim}}
We non-dimensionalize the system using  the following scaling; 
\begin{subequations}
\begin{align}
& x = \frac{ \tilde{x} }{ X_0 }, \quad t = \frac{ \tilde{t} }{ X_0^{2+3/n }/\Gamma_n}, \quad h = \frac{ \tilde{h} }{ X_0 }, \label{eq:scaling}  \\
& \eta( x ) = \frac{ \eta_0 - \tilde{ \eta } }{ \eta_0 - \eta_b } = \left[ 1 - \left\{ \frac{ d_0 }{ \lambda_n } \left( \frac{ \mathcal{G} X_0 }{ T_m } \right)^{1/n} \right\}^{\gamma} x^{\gamma/n} \right] \nonumber \\
& \qquad\qquad\qquad \equiv \left[ 1 - d_{\gamma, n} x^{\gamma/n} \right], \qquad 
\text{where} \quad
d_{\gamma, n} = \left\{ \frac{d_0}{\lambda_n}\left(\frac{\mathcal{G} X_0}{T_m}\right)^{1/n} \right\}^{\gamma}, \label{eq:dimless_visco}
\end{align}
\end{subequations}
from which we obtain 
\begin{subequations}
\begin{align}
& \partial_t h + \partial_x \left[ \frac{x^{-3/n}}{\eta_1(x)} \left( 1 + \alpha h_{xxx} \right) \right] = 0, \label{eq:mass_conservation_dimless} \\ 
& \eta_1(x) = 1 + (\eta_r - 1)d_{\gamma,n} x^{\gamma/n}, \label{eq:eta1}
\end{align}
\end{subequations}
where $\alpha = \tilde{\alpha}/X_0^2$ and $\eta_r = \eta_0/\eta_b$. 
Here, $X_0$ is a length scale chosen based on the magnitude of the temperature gradient $\mathcal{G}$, which determines the temperature drop across the size of the sample (Figure \ref{fig:schematic}).  Thus, for a sufficiently large experimental cell, the membrane height at the lowest temperatures (the cell center) will be unaffected by the deformation at the highest temperature, where $x=0$ \cite{Wilen1995, Wettlaufer1996}. This, along with appeal to experiment, leads to the following boundary conditions: 
\begin{subequations}
\begin{align}
& \partial_x h = \partial_{xx} h = 0 \qquad (x = 0), \label{eq:LBC} \qquad \textrm{and} \\
& \partial_x h = 0 \qquad (x = 1). \label{eq:RBC1} 
\end{align}
\end{subequations}
The boundary conditions
\eqref{eq:LBC} correspond to the bulk solid-liquid interface, where the slope and curvature of the membrane height vanish. 
Equation \eqref{eq:RBC1} is a ``no flux condition'' for the membrane height at the cold end, far from the warm end, but note there is {\em a fluid flux} at the cold end. The initial condition for the membrane height, which is measured relative to a reference height, is 
\begin{equation}
\label{eq:IC}
h(x, t = 0) = 0. 
\end{equation} 
To avoid a singularity at the bulk solid-liquid interface, we impose the following necessary condition on the solution
\begin{equation}
\label{eq:NC} 
\partial_{xxx} h = -\alpha^{-1} \qquad (x = 0). 
\end{equation} 

In order to obtain a unique membrane height we need to specify the fourth boundary condition. In the original theory, described in \S \ref{sec:similarity_sol} below, wherein the bulk liquid viscosity is used, the resulting family of PDEs (namely \eqref{eq:mass_conservation_dimless} with a constant viscosity) admits a similarity solution, leading to a family of ODEs and associated boundary conditions; \eqref{eq:old_similarity_ode}--\eqref{eq:similarity_necessary} \cite{Wettlaufer1996}. Analysis of the evolution equation \eqref{eq:old_similarity_ode} far from the origin provides the fourth boundary condition, \eqref{eq:similarity_decay}, which is satisfied numerically using a shooting method. Thus, in that study there was no need to solve the PDEs for $h(x,t)$, but the introduction of the proximity effect here effectively imposes an ``external'' length scale (at $x$ in the interior of the domain) that leads to a parameter dependent breakdown of the similarity solution. Hence, here we must solve the family of PDEs, for which we specify the curvature at the cold end ($x = 1$). This differs from using an asymptotic analysis to determine the boundary condition, but as discussed in Appendix \ref{sec:BC} the influence of this boundary condition is only important for parameter values for which the magnitude of the oscillations in the membrane are large at very low temperatures. The experiments done thus far \cite{Wilen1995} have been under conditions wherein the majority of the membrane deformation occurs near the warm end. Thus, at the cold end we impose 
\begin{equation}
\label{eq:RBC2}
\partial_{xx} h = 0 \qquad (x = 1). 
\end{equation}
We note, however, for analysis of future experiments across a variety of materials this condition will depend on both the control parameters as well as experimental geometry. 

Analysis of \eqref{eq:eta1} shows that the liquid viscosity across the film is dominated by the bulk value for the power law exponent $\gamma \geq \gamma_{\rm sat}$ and confinement-induced value for $\gamma \leq \gamma_{\rm crit}$, whereas the bulk and confinement-induced values are comparable for $\gamma_{\rm crit} \leq \gamma \leq \gamma_{\rm sat}$ \cite{Pramanik2017}. 
\textcolor{black}{
The dimensionless viscosity of the film is $\eta_1(x) = 1 + (\eta_r - 1)d_{\gamma,n} x^{\gamma/n} \equiv 1 + \epsilon(\gamma) x^{\gamma/n}$.
For the relevant parameter values, one can show that $\epsilon(\gamma)$ is a decreasing function of $\gamma$.  Hence, $\gamma_{crit}$ ($\gamma_{sat}$) is the largest (smallest) approximate value of $\gamma$ such that near the warm (cold) end the confinement-induced value of the viscosity of the thin liquid film, $\epsilon(\gamma) x^{\gamma/n}$, is of the same order of magnitude as that of its bulk value $\eta_b$.
See Appendix \ref{sec:gammas} for estimates of $\gamma_{crit}$ and $\gamma_{sat}$.}

We rewrite the mass conservation equation \eqref{eq:mass_conservation_dimless} as 
\begin{equation}
\label{eq:diffusion_4th_order}
\partial_t h = \mathcal{D}(x) h_{xxxx} + \mathcal{U}(x)h_{xxx} + \mathcal{F}(x), 
\end{equation}
where 
\begin{subequations}
\begin{align}
& \mathcal{D}(x) = -\alpha \frac{x^{-3/n}}{\eta_1(x)} < 0, \label{eq:D} \\
& \mathcal{U}(x) = \frac{{\rm d}\mathcal{D}}{{\rm d}x} = \alpha \frac{x^{-(1+3/n)}}{\eta_1^2(x)}\frac{(3+\gamma)\eta_1(x) - \gamma}{n} > 0, \quad\textrm{and}\label{eq:U} \\ 
& \mathcal{F}(x) = \frac{ \mathcal{U}(x) }{ \alpha } = \frac{x^{-(1+3/n)}}{\eta_1^2(x)}\frac{(3+\gamma)\eta_1(x) - \gamma}{n} > 0, \label{eq:F} 
\end{align}
\end{subequations}
which is the form solved numerically using standard second order finite difference formulae and the time integration is performed using a semi-implicit Crank-Nicolson (C-N) method \cite{LeVeque2007a}. We extend the method and stability analysis of the numerical scheme discussed in \cite{Pramanik2017} for a second order equation to the fourth order evolution equation \eqref{eq:diffusion_4th_order}. We have used a uniform grid spacing $\Delta x = x_{j+1} - x_{j}$, $\forall j \in \{ -2, -1 , 0, \hdots$ $,N, N+1, N+2 \}$ for the spatial variable, such that
\[ 
x_{-2} < x_{-1} < (0 =) x_0 < x_1 < \hdots < x_{N-1} < x_{N} (= 1) < x_{N+1} < x_{N+2},  
\]
and the time stepping is so chosen that the numerical stability is met \cite{Pramanik2017}. (See Appendix \ref{sec:code_valid} for grid independence test.) 



\section{Similarity solutions\label{sec:similarity_sol}} 
For the bulk viscosity case, $\eta_r = 1$, a family of similarity solutions, described by a single parameter $n$, is obtained as
\begin{subequations}
\begin{align}
& g(\zeta) = \tilde{h} \tilde{\alpha} \left( \Gamma_n \tilde{\alpha} \tilde{t} \right)^{-3n/(4n+3)}, \label{eq:old_similarity_sol} \\
& \zeta = \tilde{x} \left( \Gamma_n \tilde{\alpha} \tilde{t} \right)^{-n/(4n+3)},  \label{eq:old_similarity_var}
\end{align}
\end{subequations}
that satisfies the following fourth order family of ordinary differential equations \cite{Wettlaufer1996}: 
\begin{subequations}
\begin{align}
g^{\prime \prime \prime \prime} &= \frac{3}{n}\frac{g^{\prime \prime \prime} + 1}{\zeta} + \frac{n}{4n+3}\zeta^{(3/n)+1}g^{\prime} - \frac{3n}{4n+3}\zeta^{3/n}g, \label{eq:old_similarity_ode} \\
 g^{\prime} &= g^{\prime \prime} = 0 \qquad (\zeta = 0), \label{eq:similarity_LBC} \\ 
g^{\prime} &= 0 \qquad (\zeta = \zeta_{\rm c}), \label{eq:similarity_RBC} \\ 
g &\sim \zeta^{-[(3/n)+1]} \qquad (\zeta \to \infty), \quad\textrm{and}\label{eq:similarity_decay} \\
g^{\prime \prime \prime} &= -1 \qquad (\zeta = 0), \label{eq:similarity_necessary}
\end{align}
\end{subequations}
where, $\zeta_c$ denotes the cell center. This similarity solution is valid for both $\eta_r = 1$ \cite{Wettlaufer1996} and for parameter dependent values of the viscosity exponent from \eqref{eq:dimless_visco}, $\gamma \geq \gamma_{\rm sat}$,  heuristically like the case for a second order system that treats a related problem in a different geometry \cite{Pramanik2017}.  

On the other hand, when the viscosity exponent is $\gamma \leq \gamma_{\rm crit}$ \cite{Pramanik2017}, we obtain a two parameter ($n, \; \gamma$) family of similarity solutions,  
\begin{equation}
\label{eq:new_similarity_sol}
g_{\gamma}(\zeta_{\gamma}) = \tilde{h} \tilde{\alpha} \left[ \frac{ \Gamma_n \tilde{\alpha} \tilde{t} }{ \delta } \right]^{-3n/(4n+3+\gamma)}, 
\end{equation}
with the similarity variable 
\begin{equation}
\label{eq:new_similarity_var}
\zeta_{\gamma} = \tilde{x} \left[ \frac{ \Gamma_n \tilde{\alpha} \tilde{t} }{ \delta } \right]^{-n/(4n+3+\gamma)},  
\end{equation}
where 
\begin{equation}
\label{eq:delta}
\delta = \left( \eta_r - 1 \right) \left[ \frac{d_0}{\lambda_n} \left( \frac{\mathcal{G}}{T_m} \right)^{1/n} \right]^{\gamma}, \nonumber 
\end{equation}
that are the solutions to 
\begin{eqnarray}
\label{eq:new_similarity_ode}
& & g_{\gamma}^{\prime \prime \prime \prime} = \frac{3+\gamma}{n}\frac{g_{\gamma}^{\prime \prime \prime} + 1}{\zeta_{\gamma}} + \frac{n}{4n+3+\gamma}\zeta_{\gamma}^{(3+n+\gamma)/n}g_{\gamma}^{\prime} - \frac{3n}{4n+3+\gamma}\zeta_{\gamma}^{(3+\gamma)/n}g_{\gamma}, 
\end{eqnarray}
subject to the conditions, \eqref{eq:similarity_LBC}, \eqref{eq:similarity_RBC}, and \eqref{eq:similarity_necessary}, written in terms of the appropriate dependent ($g_{\gamma}$) and independent ($\zeta_{\gamma}$) variables. As $\zeta_{\gamma} \to \infty$, asymptotic analysis of 
\eqref{eq:new_similarity_ode} reveals that the membrane height decays algebraically as 
\begin{equation}
\label{eq:new_similarity_decay}
g_{\gamma} \sim \zeta_{\gamma}^{-(3+n+\gamma)/n} \qquad (\zeta_{\gamma} \to \infty). 
\end{equation}

\begin{figure}[hbtp]
\centering
(a) \hspace{2.4 in} (b) \\ 
\includegraphics[scale=0.5]{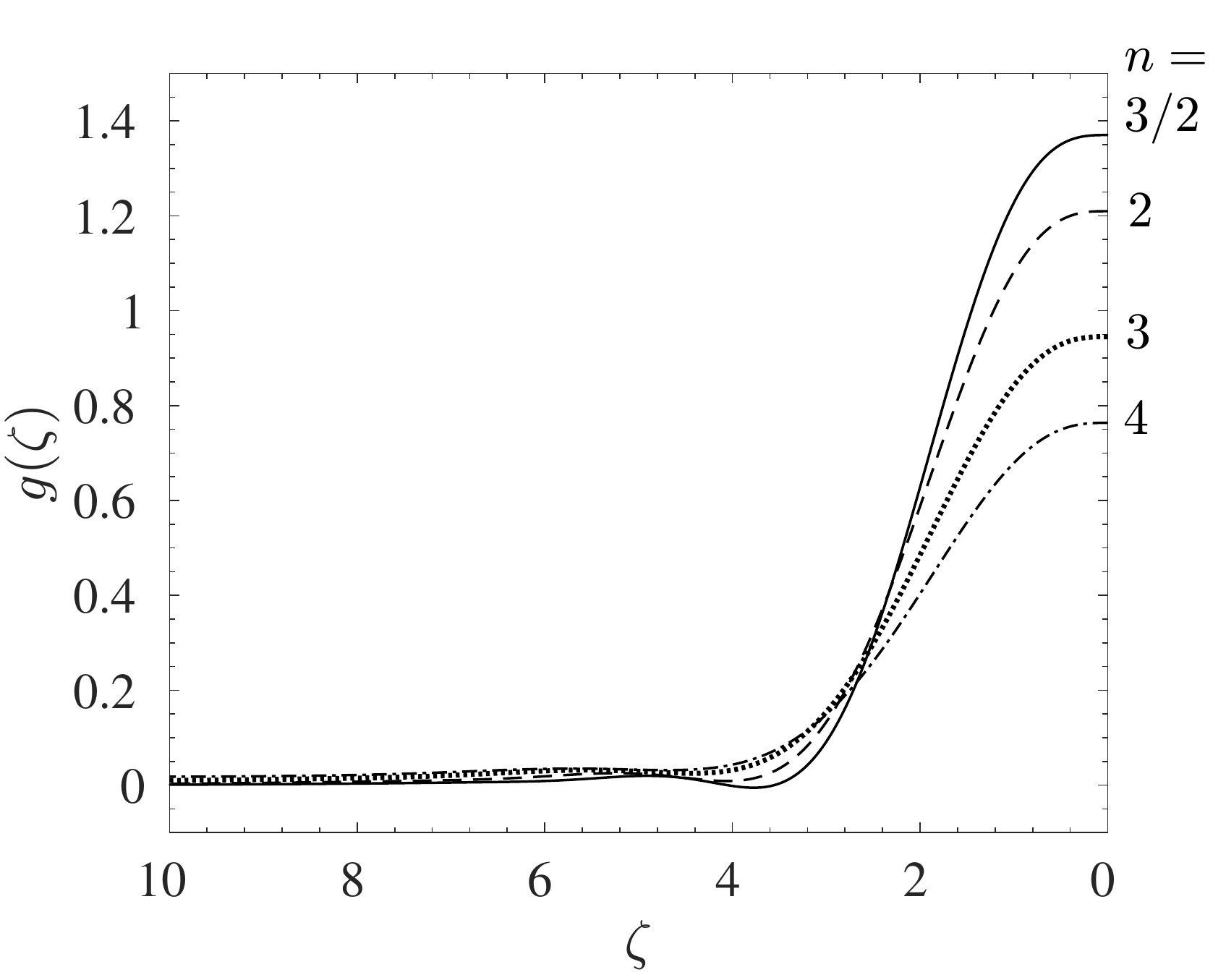} 
\includegraphics[scale=0.5]{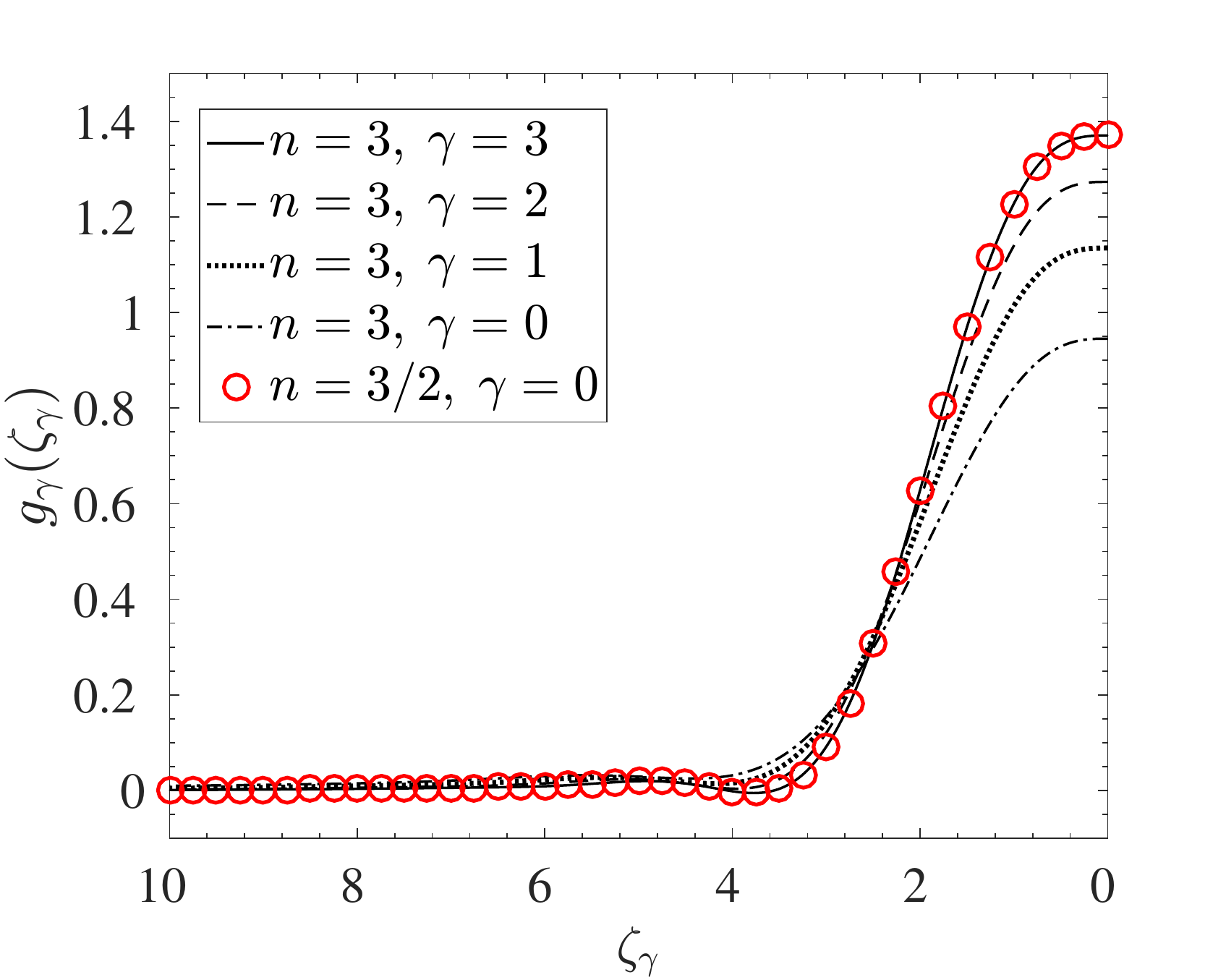}
\caption{(a) Similarity solutions of the system \eqref{eq:old_similarity_ode}-\eqref{eq:similarity_necessary}, reproducing figure 2 of \cite{Wettlaufer1996}. (b) Similarity solutions of \eqref{eq:new_similarity_ode} for non-retarded van der Waals interactions ($n = 3$) and four values of $\gamma = 0, \; 1, \; 2, \; 3$. Also shown is the similarity solution $g_0(\zeta_0)$ for $n = 3/2$, $\lambda_{n = 3/2} = 0.0337$ \AA \cite{Wettlaufer1996}. From the scaling of the similarity variable, $g_{\gamma}(\zeta_{\gamma})$ we see that $g_{0}[\zeta_0(n = 3/2)] \equiv g_{3}[\zeta_3(n = 3)]$, a condition successfully captured by our numerical solutions.}
\label{fig:validation}
\end{figure}

The system \eqref{eq:old_similarity_ode}-\eqref{eq:similarity_necessary} (or, the version corresponding to the two parameter family of similarity solutions $g_{\gamma}(\zeta_{\gamma})$) is solved numerically using a finite difference method with uniform mesh. To validate our numerical method, we reproduce figure 2 of \cite{Wettlaufer1996}, shown here in Figure \ref{fig:validation}(a). Figure \ref{fig:validation}(b) compares five members of the second family of similarity solutions, $g_{\gamma}(\zeta_{\gamma})$. 

\section{Results\label{sec:Results}} 
Here, we investigate the time evolution of the membrane height, described by the family of fourth order parabolic differential equations \eqref{eq:mass_conservation_dimless}, whose solutions exhibit oscillatory behavior as is common in a wide class of fourth order equations appearing in many settings (See e.g., \cite{Oron:1997, Bertozzi:1998, Craster:2009, Chapman:2013fk}, and Refs. therein).
In this section we present the numerical results for (a) temperature gradients of $\mathcal{G} = 0.92 \times 10, ~ 0.92 \times 10^2$ and $0.92 \times 10^3$ K/m, (b) molecular diameters of $d_0 = 1, ~ 3, ~ 5, ~ 8$ and 10 \AA, and (c) a wide range of the power law exponent $\gamma$ in connection with $\eta_r$. Unless otherwise stated, the other physical parameters used are listed in Table \ref{tab:parameters}. 

\begin{table}[!h]
\caption{List of parameters}
\label{tab:parameters}
\centering
\begin{tabular}{|l|l|l|l|}
\hline
Parameters & Values & Units \\
\hline
	$\rho_l$ &  $10^3$ &  ${\rm kg/m}^{3}$ \\
	$\rho_s$ & $0.9167 \times 10^3$ & ${\rm kg/m}^3$ \\ 
	$T_m$ & $273.16$ & K \\ 
	$\eta_b$ & $1.307 \times 10^{-3}$ &  kg/m-s \\ 
	$q_m$ & $334 \times 10^3$ & J/kg \\ 
	$\tilde{\sigma}$ & 1 & psi-cm \\ 
	$(n, \lambda_n)$ & $(3, 1.3759)$ & (-, \AA) \\ 
	$(\mathcal{G}, \; X_0)$ & $(0.92 \times 10, \; 2 \times 10^{-3})$ & (K/m, m) \\ 
	$(\mathcal{G}, \; X_0)$ & $(0.92 \times 10^2, \; 10^{-3})$ & (K/m, m) \\ 
	$(\mathcal{G}, \; X_0)$ & $(0.92 \times 10^3, \; 10^{-3})$ & (K/m, m) \\ \hline 
\end{tabular}
\vspace*{-4pt}
\end{table}

\begin{figure}[hbtp]
\centering
(a) \hspace{2.4 in} (b) \\ 
\includegraphics[scale=0.5]{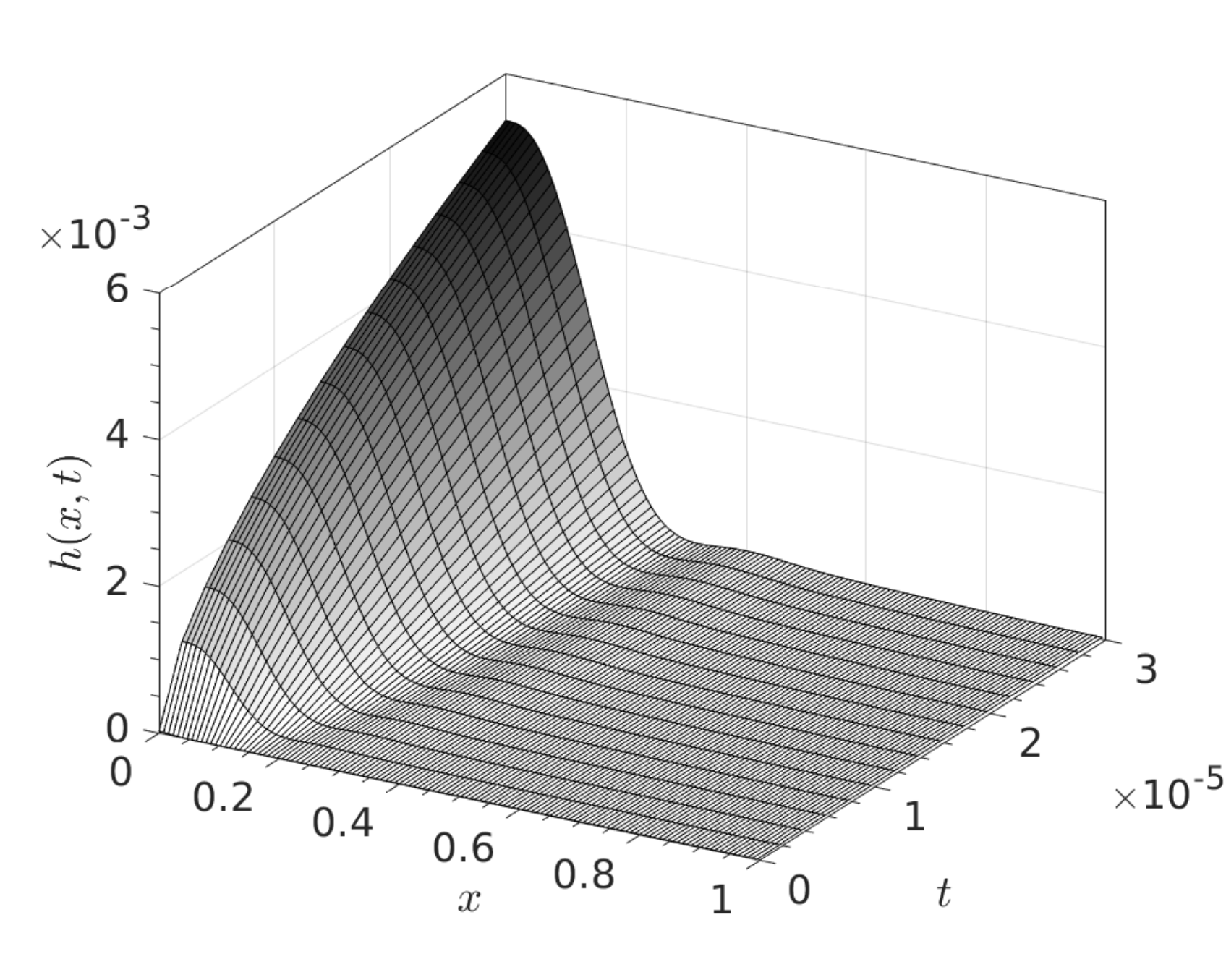} 
\includegraphics[scale=0.5]{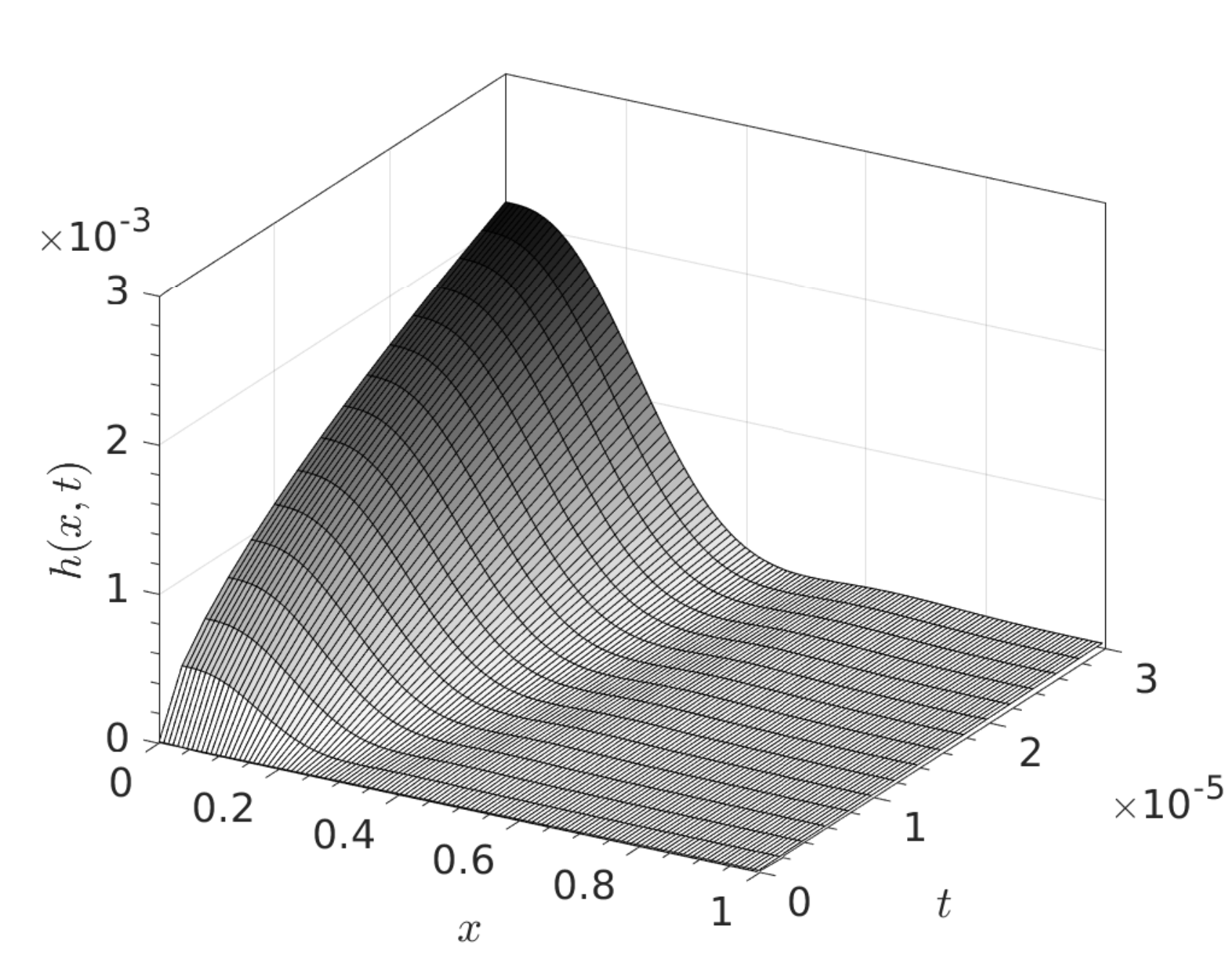} 
\caption{Evolution of the dimensionless membrane height for the bulk viscosity model for (a) a large temperature gradient, 
$\alpha = 0.0669$, 
and (b) a small temperature gradient, 
$\alpha = 0.6686$.
Note the different scales of the axes.} 
\label{fig:effect_G}
\end{figure}

\subsection{Effects of the temperature gradient $ \mathcal{G}$} 
\textcolor{black}{
Here we discuss the effect of the temperature gradient on the evolution of the membrane height. Note that both the dimensionless parameters $\alpha$ and  $d_{\gamma, n}$ depend on $\mathcal{G}$. For the case of bulk viscosity model, premelting dynamics is independent of $d_{\gamma, n}$; the only dimensionless parameter that captures the effect of the temperature gradient is $\alpha \propto \mathcal{G}^{-1}$. Figure \ref{fig:effect_G} shows the membrane height, $h(x, t)$, for the bulk viscosity case and a large ($\alpha = 0.0669$) and small ($\alpha = 0.6686$) temperature gradient.  As $\alpha$ decreases, deformation is localized near the warm end, but the qualitative properties of the deformation are unaffected. Namely, as $\alpha$ decreases, because $\tilde{d}^3 \propto t_r^{-3/n} \propto \left( \alpha^{-1} {x} \right)^{-3/n}$, the film thins rapidly towards the lower temperatures. For example, in the case of nonretarded van der Waals interactions ($n = 3$), a decrease in $\alpha$ by a factor of ten, reduces the film thickness more than $50\%$ everywhere in the domain. Thus, mass conservation dictates that more premelted liquid freezes at higher temperatures, as seen in Figure \ref{fig:effect_G}.
Similar qualitative effects of a steady temperature gradient on the deformation of an elastic capillary wall with a local hoop stress are seen \cite{Wettlaufer1995, Pramanik2017}, because the main physical effects are at play; as the temperature gradient increases, the gradient in the liquid volume flux is localized at higher temperatures.}

\begin{figure}[hbtp]
\centering
(a) \hspace{2.4 in} (b) \\
\includegraphics[scale=0.5]{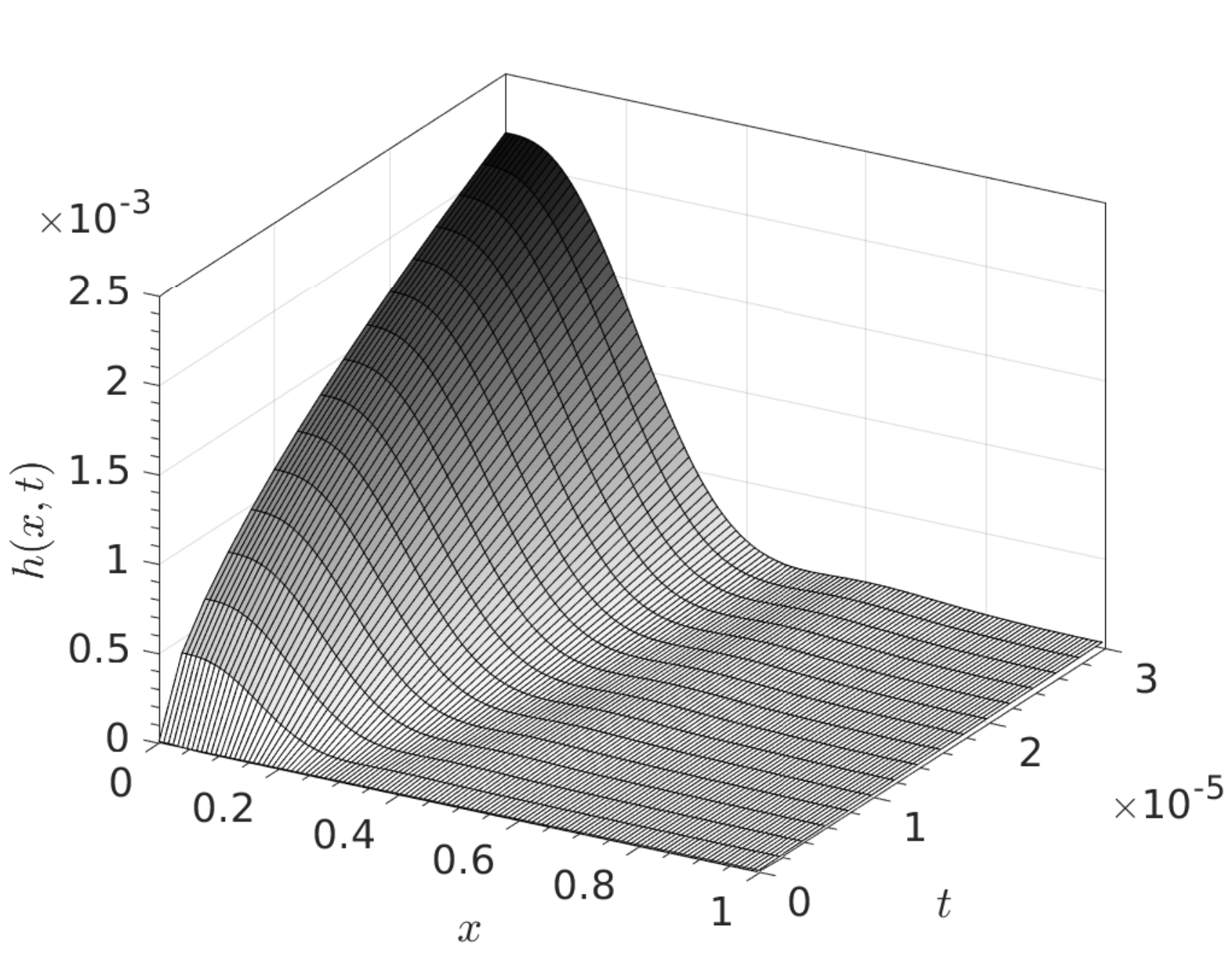} 
\includegraphics[scale=0.5]{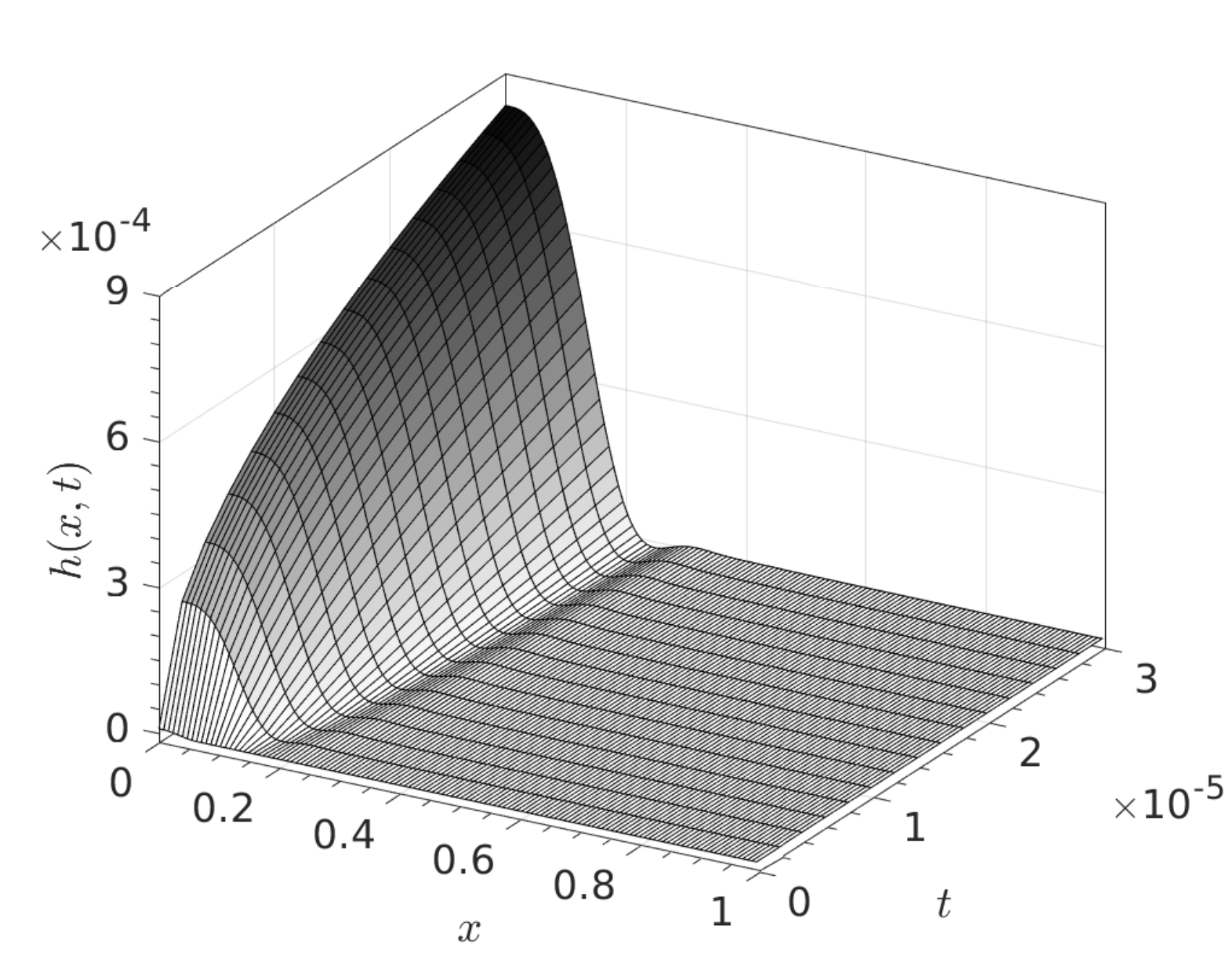} 
\caption{Evolution of the dimensionless membrane height for 
$\eta_r = 10^5$, $\alpha = 0.6686$, 
(a) $d_{6, 3} = 1.2188 \times 10^{-5}$, and (b) $d_{6, 3} = 1.6719 \times 10^{-2}$. Note the different scales of the axes.
} 
\label{fig:effect_d0}
\end{figure}

\subsection{Effects of short range cut-off, $d_0$} 
We of course know the size of a molecule, but depending on the crystallographic orientation, the ordering effect on the liquid differs \cite{Nada:1997, Nada:2000, Nada:2016}.  Hence, we view the short range cut-off, $d_0$, as a proxy for crystallographic orientation and its ordering influence. Recall from \eqref{eq:dimless_visco} that 
\begin{equation}
\label{eq:d_gamma_n}
d_{\gamma, n} = \left\{ \frac{d_0}{\lambda_n}\left(\frac{\mathcal{G} X_0}{T_m}\right)^{1/n} \right\}^{\gamma}, 
\end{equation}
and hence for fixed values of $\lambda_n, ~ \mathcal{G}, ~ X_0, ~ T_m$ and the exponents $n$ and $\gamma$, we see that as $d_0$ increases both $d_{\gamma, n}$ and $\eta_1(x)$ increase (see Eq. \eqref{eq:eta1}). Thus, viscous resistance (volume flux) in the thin liquid film increases (decreases) more rapidly parallel to the temperature gradient as $d_0$, and hence the ordering influence, increases.  Therefore, the solidification and concomitant deformation, is localized near the warm end, qualitatively like increasing the temperature gradient. For the bulk viscosity model ($\eta_r = 1$) premelting  dynamics is independent of $d_0$. However, $d_0$ has a significant influence on the solidification in the proximity viscosity model. 
\textcolor{black}{
Figure \ref{fig:effect_d0} shows the membrane evolution, $h(x, t)$, for $\eta_r = 10^5$, $\alpha = 0.6686$, and different values of $d_0$. For each $d_0$, we compute $\gamma_{\rm crit}$, $\gamma_{\rm sat}$ and use $\gamma_{\rm crit} < \gamma < \gamma_{\rm sat}$ in figure \ref{fig:effect_d0}. The corresponding values of the dimensionless parameter $d_{\gamma, n}$ are $d_{6, 3} = 1.2188 \times 10^{-5}$, $1.6719 \times 10^{-2}$.} 

\begin{figure}[hbtp]
\centering
(a) \hspace{2.4 in} (b) \\ 
\includegraphics[scale=0.5]{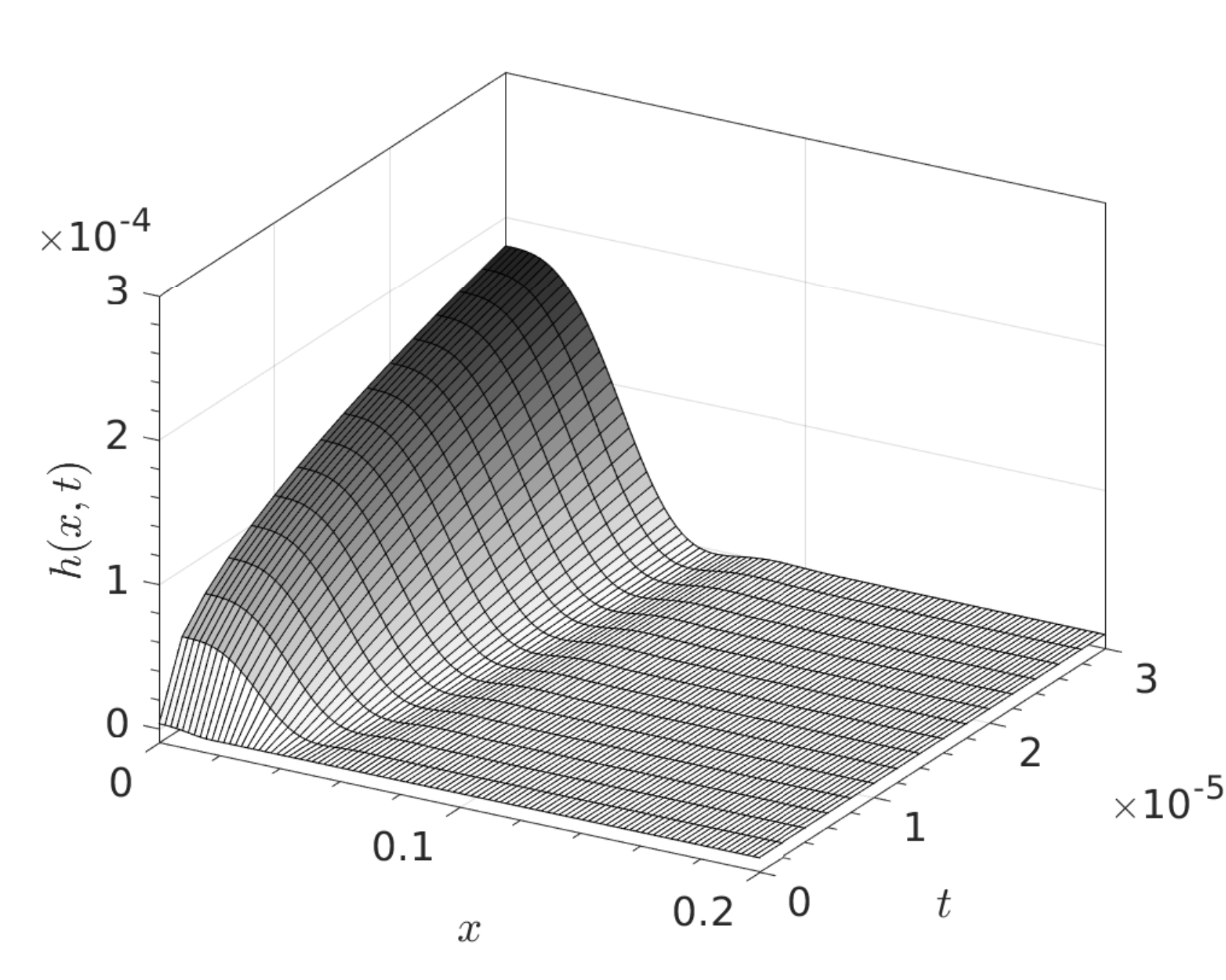} 
\includegraphics[scale=0.5]{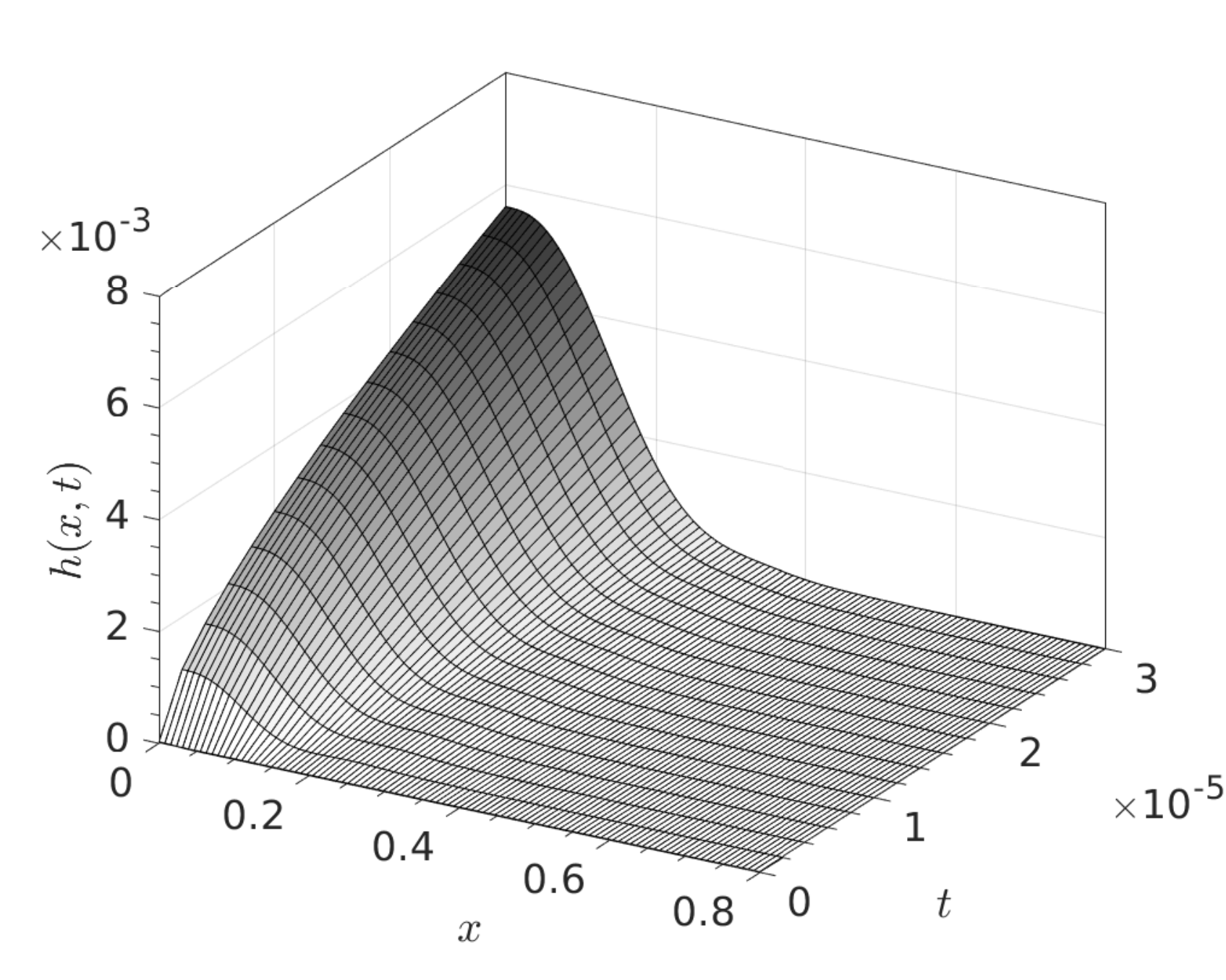}
\caption{Membrane height evolution for 
$\eta_r = 10^7$, $\alpha = 0.0669$ and (a) $d_{5.1, 3} = 4.51 \times 10^{-2}$, and (b) $d_{15, 3} = 1.1031 \times 10^{-4}$.
Note the differences in the scale of the axes.} 
\label{fig:surface1_etar1e7}
\end{figure}

\subsection{The influence of the proximity effect} 
In this section we discuss the influence of the power law exponent, $\gamma$, which governs the strength of the decay of the proximity effect, on the evolution of the membrane. We vary $\gamma$ over a wide range and study the numerical solutions for $\eta_r = 10^3, 10^5$ and $10^7$. We recover the similarity solutions \eqref{eq:old_similarity_sol} and \eqref{eq:new_similarity_sol} when $\gamma \leq \gamma_{\rm crit}$ and $\gamma \geq \gamma_{\rm sat}$, respectively. \textcolor{black}{For a given value of $\alpha$ this effect is captured via $d_{\gamma, n}$. Here we consider $\alpha = 0.0669$ and non retarded van der Waals interaction ($n = 3$).}

In the bulk viscosity model the principal positive curvature high temperature maximum leads to an oscillation as follows. Firstly, membrane elasticity acts to suppress the curvature, the response to which is to draw in more premelted fluid. Secondly, the fluid drawn in from the low temperature region creates a negative curvature that elasticity acts to suppress driving fluid towards both the low and high temperature directions. Finally, the fluid driven towards the low temperature direction creates a new peak, and the process repeats itself \cite{Wettlaufer1996}. Here, the same basic oscillation in the membrane height appears when $\gamma \leq \gamma_{\rm crit}$, and so we focus  on the membrane evolution for $\gamma \in (\gamma_{\rm crit}, \gamma_{\rm sat})$. 

\begin{figure}[hbtp]
\centering
\includegraphics[scale=0.5]{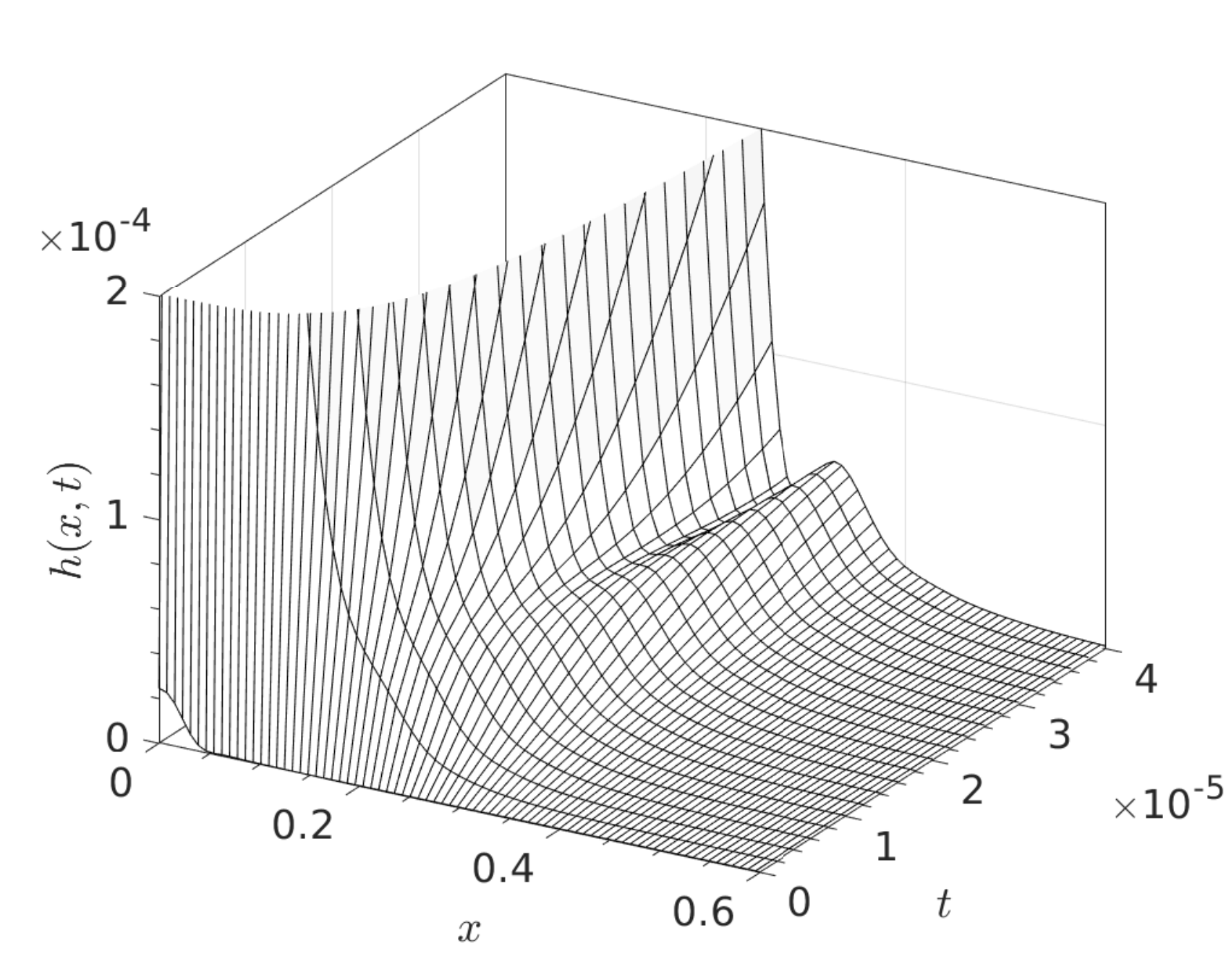} 
\caption{Membrane evolution between $h = 0$ and $h = 2 \times 10^{4}$, which surround the first minimum and the second maximum, for 
$\eta_r = 10^7$, $\alpha = 0.0669$, $d_{13, 3} = 3.7175 \times 10^{-4}$.
} 
\label{fig:surface2_etar1e7}
\end{figure}

\textcolor{black}{
In Figures \ref{fig:surface1_etar1e7} we show the membrane evolution for $\eta_r = 10^7$, $\alpha = 0.0669$ and $d_{\gamma, 3} = 4.51 \times 10^{-2} \; (\gamma = 5.1)$, $1.1031 \times 10^{-4} \; (\gamma = 15)$. Comparing Figure \ref{fig:effect_G}(a) with Figure \ref{fig:surface1_etar1e7}(a) we see that the maximum deformation of the membrane is one order of magnitude larger for the bulk viscosity case than that for $\gamma = 5.1 \approx \gamma_{\rm crit}$. On the other hand, for $\gamma = 19 < \gamma_{\rm sat}$, the maximum deformation is nearly the same as that for the bulk viscosity model, which is consistent with the analysis used to predict $\gamma_{\rm sat}$ for a similar problem \cite{Pramanik2017}. However, for $\gamma = 15$, in Figure \ref{fig:surface1_etar1e7}(b) we observe two differences from the bulk viscosity model. Firstly, the maximum membrane deformation $h_{\rm max}$ is larger than in the bulk viscosity case and secondly, membrane oscillations are suppressed. The interaction between the elasticity induced curvature and the proximity effect on the dynamics of the membrane height can be seen by examining the membrane evolution in a region surrounding the first minimum and the second maximum.  This is shown for $d_{\gamma, 3} = 3.7175 \times 10^{-4} \; (\gamma = 13)$ in Figure \ref{fig:surface2_etar1e7}, with the other parameter values the same as in Figure \ref{fig:surface1_etar1e7}. It is seen that the oscillations vanish for a parameter dependent time interval; the same general behavior is found for a wide range of temperature gradients ($\alpha$), molecular diameters ($d_0$) and viscosity ratios ($\eta_r$), as discussed for $\gamma \in (\gamma_{\rm crit}, \gamma_{\rm sat})$ in Section \ref{subsec:wiggle} and Appendix \ref{sec:additional}. 
}


%
 
\subsubsection{Reentrant oscillations of $h$\label{subsec:wiggle}} 
As discussed in section \ref{sec:similarity_sol}, the existence of an infinite number of peaks in the membrane height with an exponentially decaying amplitude is found for the case of bulk viscosity (e.g., Figure \ref{fig:validation}), and such oscillations are influenced by the confinement effects modeled by \eqref{eq:powerlaw_visco_d}. 
Thus, when the confinement-induced and bulk viscosity are comparable, the associated internal length scale imposed on the system influences the membrane oscillations (Figures \ref{fig:surface1_etar1e7}(b) and \ref{fig:surface2_etar1e7}). 

For such oscillations to persist, the membrane curvature must change sign, and as seen in Figures \ref{fig:surface1_etar1e7}(b) and \ref{fig:surface2_etar1e7}, the film thickness dependent viscosity is responsible for the vanishing of the first trough and hence the oscillations. The evolution equation \eqref{eq:mass_conservation_dimless} shows how the flux operator, written as $\mathcal{D}(x)$ in \eqref{eq:D}, controls the dynamics of the membrane height. The viscosity model (Eq. \eqref{eq:eta1}) shows that the proximity effects, $\eta_1 = 1 + (\eta_r - 1) d_{\gamma, n} x^{\gamma/n} \sim (\eta_r - 1) d_{\gamma, n} x^{\gamma/n}$, dominate near the cold end ($x = 1$), whereas the bulk viscosity effects, $\eta_1 = 1 + (\eta_r - 1) d_{\gamma, n} x^{\gamma/n} \sim 1$, dominate near the warm end ($x = 0$). In terms of the flux operator, at the cold end we have $ -\mathcal{D}(x) \sim x^{-(3+\gamma)/n} $ and at the warm end we have $ -\mathcal{D}(x) \sim x^{-3/n} $, and thus the similarity solution in the bulk viscosity region grows faster, as $h \sim t^{3n/(4n + 3)}$, than the solution in the proximity-viscosity dominated region, for which $h \sim t^{3n/(4n + 3 + \gamma)}$. In the crossover between these two scaling behaviors  the membrane height dynamics is intermediate and depends on $\eta_r$ and $d_{\gamma, n}$ (i.e., $\gamma, \; n, \; \lambda_n, \; d_0, \; \mathcal{G}, \; T_m$). Namely, for $\gamma \in (\gamma_{\rm crit}, \gamma_{\rm sat})$, $-\mathcal{D}(x)$ exhibits both $\sim x^{-3/n}$ and $\sim x^{-(3+\gamma)/n}$ scalings locally within the spatial domain, giving different temporal scalings for the evolution of $h(x,t)$. 

Consider the situation wherein the first minimum of $h(x,t)$ is located in the bulk viscosity-dominated region, but the second maximum is located within the crossover region or beyond. In such a situation, the first minimum grows faster than the second maximum, which is overcome by the minimum until at a time $t^{\dagger}$, $h(x, t^{\dagger})$ decreases monotonically with $x$. At later times, when the deformation of the membrane enters the proximity viscosity-dominated region, the oscillatory structure in $h(x, t)$ reappears. When circumstances are such that there is a rapid change in the viscosity across the first minimum the fluid viscosity increases rapidly with $x$.  However, as elasticity reduces the positive curvature created by the primary maximum, it draws fluid in from both the high and low temperatures regions. However, because the viscous friction in the proximity dominated low temperature region is larger than that in the bulk viscosity high temperature region, more fluid is drawn from the higher temperature lower $x$ region. Therefore, the magnitude of the negative curvature in the low temperatures end is reduced, suppressing the the first minimum. Eventually, the primary peak spans both the bulk viscosity and crossover regions and continued oscillations can be described along similar lines.  

\begin{figure}[hbtp]
\centering
(a) \hspace{2 in} (b) \\ 
\includegraphics[scale=0.5]{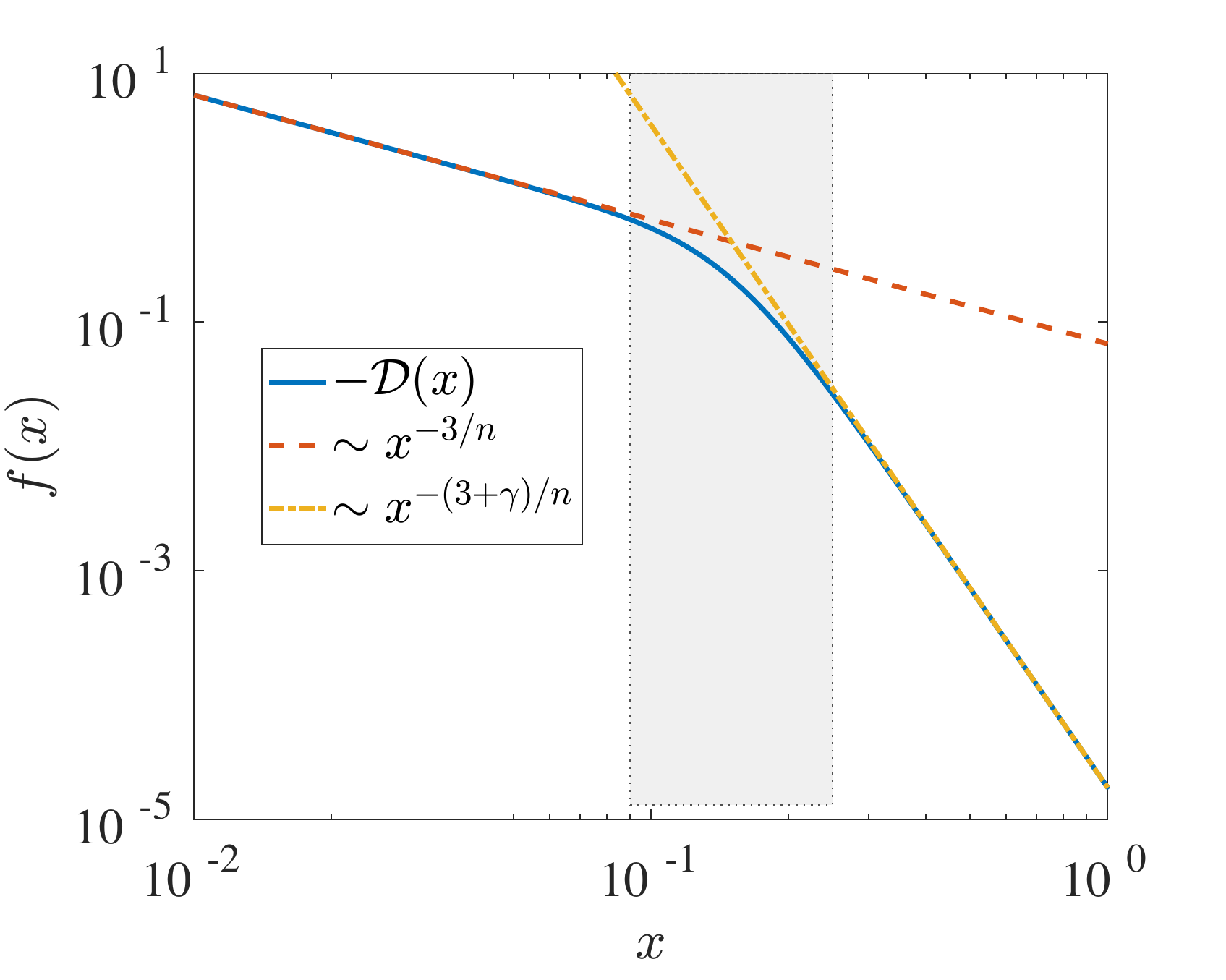} 
\includegraphics[scale=0.5]{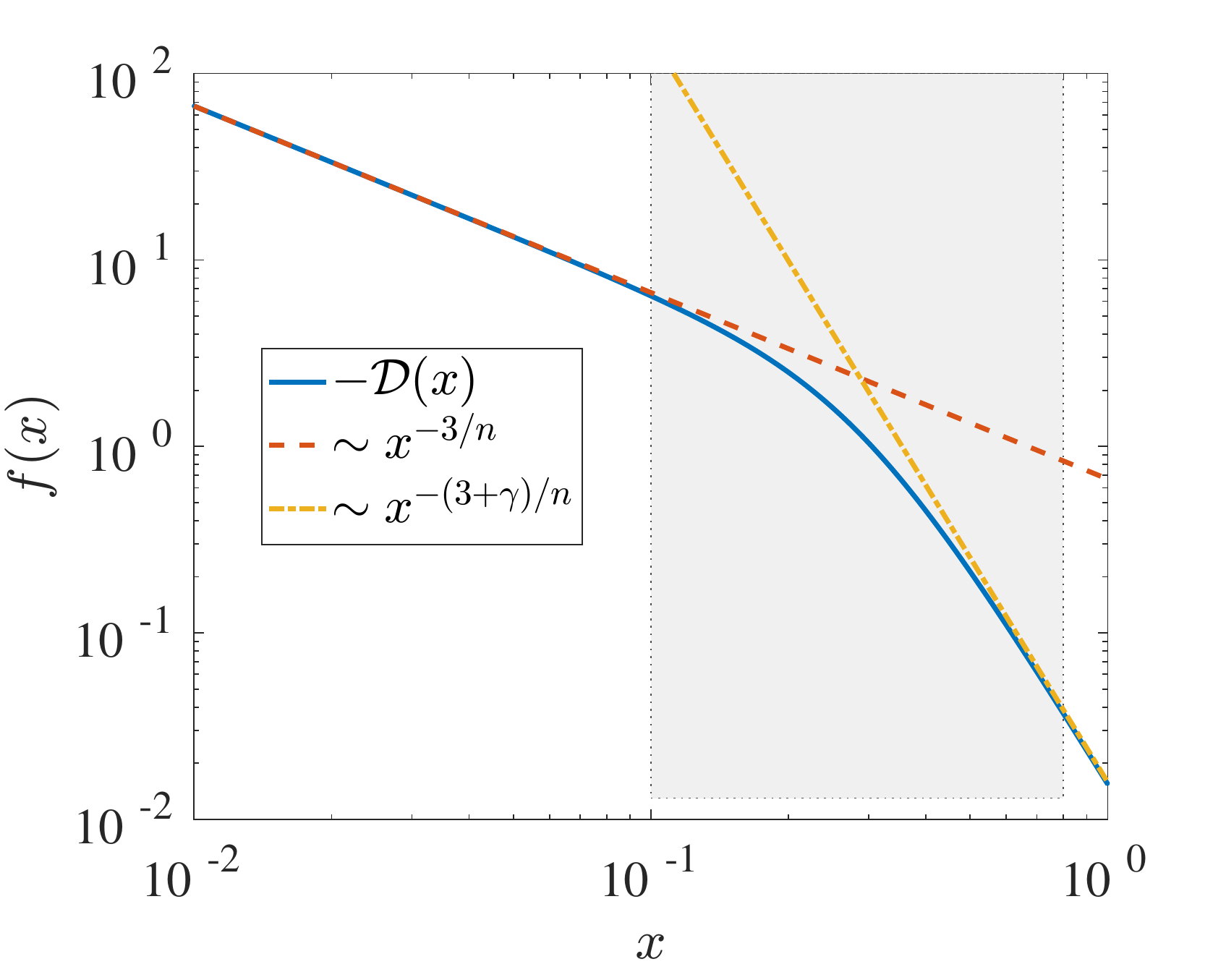} \\ 
(c) \\ 
\includegraphics[scale=0.5]{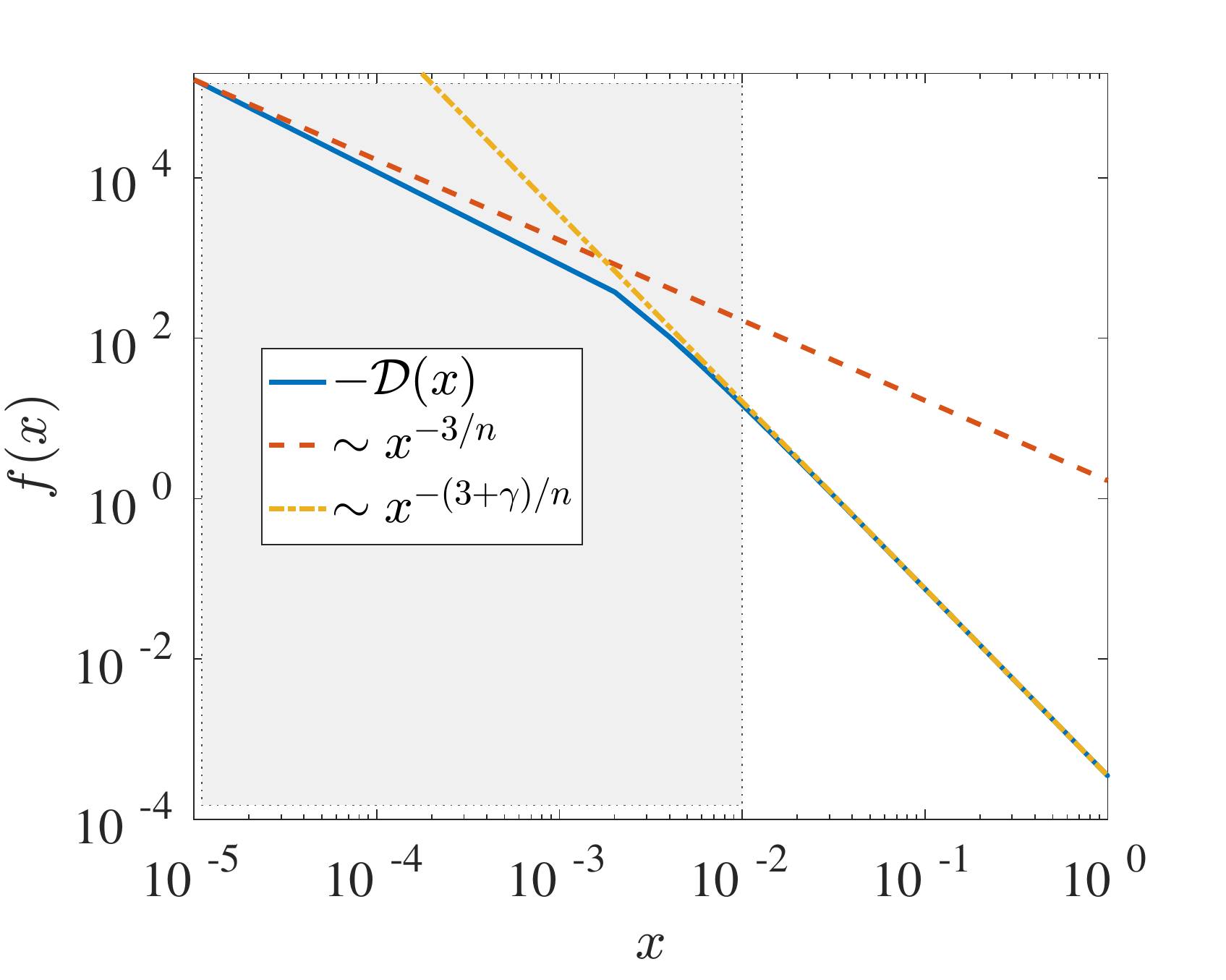} 
\caption{The ordinate is generically labelled $f(x)$ to indicate scaling behavior of $- \mathcal{D}(x) =  \displaystyle \alpha { x^{-3/n} }\left[1 + d_{\gamma, n} (\eta_r - 1) x^{\gamma/n} \right]^{-1}$, which is compared to $x^{-3/n}$ and $x^{-(3 + \gamma)/n}$ for 
$\eta_r = 10^7$. The shaded area denotes the crossover region between the two scalings. (a) For $\alpha = 0.0669$ and $d_{13, 3} = 3.7175 \times 10^{-4}$, the crossover region is spanned over $0.09 \lesssim x \lesssim 0.25$. (b) For $\alpha = 0.6686$ and $d_{9, 3} = 4.2224 \times 10^{-6}$, the crossover region is spanned over $0.1 \lesssim x \lesssim 0.8$. (c) For $\alpha = 6.6861$ and $d_{4, 3} = 1.8968 \times 10^{-4}$, the crossover region is restricted to within $x \lesssim 0.01$.} 
\label{fig:D_vs_x}
\end{figure}

\section{Discussion and Conclusions\label{sec:summary}} 
The key dynamics in the class of conservation laws that our study is a part of are embodied in the structure of the flux term.  The key distinctions between our work and the majority of such conservation laws governed by one or another manifestation of lubrication theory are that (1) our physical system has a thin film whose thickness that is a known function of position, but the evolution equation is for the elastic surface with which it is in contact and (2) we deal with two families of conservation laws, depending on the nature of the interactions controlling the film thickness and those influencing the ordering of the film as it thins.   

The structure of the flux is given by $\mathcal{D}(x)$, as expressed by \eqref{eq:D}, and is plotted in Figure \ref{fig:D_vs_x} along with the limits for (i) the bulk dominated viscosity, $\propto x^{-3/n}$, and (ii) the proximity dominated viscosity, $\propto x^{-(3 + \gamma)/n}$. \textcolor{black}{We examine $\mathcal{D}(x)$ with $\eta_r = 10^7$, and small,  intermediate, and large temperature gradients in Figures \ref{fig:D_vs_x}(a)--(c) for different values of $d_{\gamma, n}$. For example, for $\alpha = 0.0669$ and $d_{\gamma, n} = 3.7175 \times 10^{-4} \; (\gamma = 13, n = 3)$ we have
\begin{eqnarray}
\label{eq:lf_decay}
& & - \mathcal{D}(x) \sim x^{-3/n}, \quad \text{for} \; 0 \leq x \lesssim 0.09 \; (\equiv x_l), \\ 
\label{eq:rh_decay}
& & - \mathcal{D}(x) \sim x^{-(3+\gamma)/n}, \quad \text{for} \; (x_r \equiv) \; 0.25 \lesssim x \leq 1, 
\end{eqnarray}
and for the intermediate range, $x_l < x < x_r$, the flux decays with $x$ (see Figure \ref{fig:D_vs_x}(a))}. Based upon 
these flux scaling behaviors, combined with the results from Figures \ref{fig:surface1_etar1e7} and \ref{fig:surface2_etar1e7} (also Figure \ref{fig:surface3}), we make the following observations. 

Firstly, for two points $x_1$ and $x_2$ such that $x_1 \in (0, x_l)$ and $x_2 \in (x_l, x_r)$, $h(x_1, t)$ grows faster than $h(x_2, t)$. Now, if $h_1^0 \equiv h(x_1, t_0) < h(x_2, t_0) \equiv h_2^0$ at time $t = t_0$, then the differential growth of $h(x,t)$ at these two points at some later time $t_f > t_0$ is such that $h_1^f = h(x_1, t_f) > h(x_2, t_f) = h_2^f$. This also holds true for $x_3 \in (x_l, x_r)$ and $x_4 \in (x_r, 1)$. 

Secondly, when the primary high temperature maximum of $h$ spreads over $0 \leq x \leq x^{\dagger}$, where $x^{\dagger} > x_r$, then, provided $x^{\dagger}$ is at a sufficiently warm position, the second crest and first trough reappear. 

Thirdly, there are times $t_1$ and $t_2$ such that $\forall t \in (t_1, t_2)$, $h(x, t)$ is a monotonic function of $x$.  (We have not obtained analytical approximations of $t_1$ and $t_2$.) For fixed values of 
\textcolor{black}{$\eta_r$ and $\alpha$,}
the crossover region moves towards the cold (warm) end as $\gamma$ increases (decreases) in the interval $(\gamma_{\rm crit}, \gamma_{\rm sat})$. Therefore, both $t_1$ and $t_2$ increase with $\gamma \in (\gamma_{\rm crit}, \gamma_{\rm sat})$. 

\textcolor{black}{
For example, Figure \ref{fig:D_vs_x}(b) shows that the crossover region covers almost $70\%$ of the domain for $\alpha = 0.6686$ and $d_{9, 3} = 4.2224 \times 10^{-6}$. Therefore, in this case we anticipate a large time window $(t_1, t_2)$. Whereas for $\alpha = 6.6861$ and $d_{7, 3} = 3.0658 \times 10^{-7}$, the crossover region spans more than $80\%$ of the domain. In this case, a second crest located within the crossover region reappears for a short intermediate time interval, showing that the existence and the length of the time interval for $h(x, t)$ to be a monotonic function of $x$ depends nonlinearly on $d_{\gamma, n}$. On the other hand, for $\alpha = 6.6861$ and $d_{4, 3} = 1.8968 \times 10^{-4}$ (and the same parameters as above), the crossover region is confined to a boundary layer near the warm end ($x \lesssim 0.01$, Figure \ref{fig:D_vs_x}(c)). Therefore, we expect oscillations in $h(x, t)$ for ostensibly all $t > 0$. (Note that $( t_2 - t_1 )$ should be compared for different $\gamma \in (\gamma_{\rm crit}, \gamma_{\rm sat})$ for fixed values of $\alpha$, $\eta_r$, and $n$ as the prefactor of the similarity solutions depends on all of these parameters.) 
}

\begin{figure}[hbtp]
\centering
(a) \hspace{2.4 in} (b) \\ 
\includegraphics[scale=0.5]{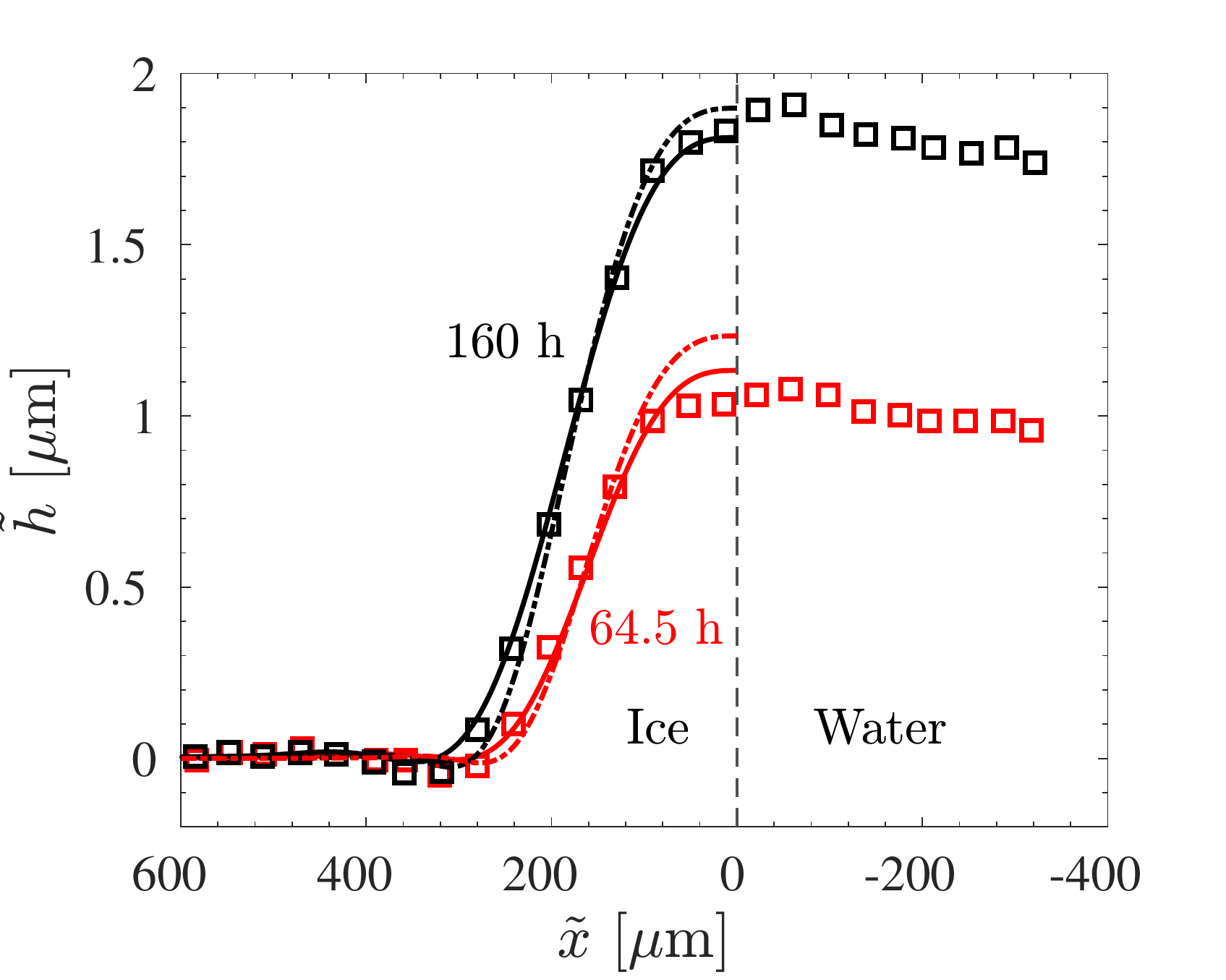} 
\includegraphics[scale=0.5]{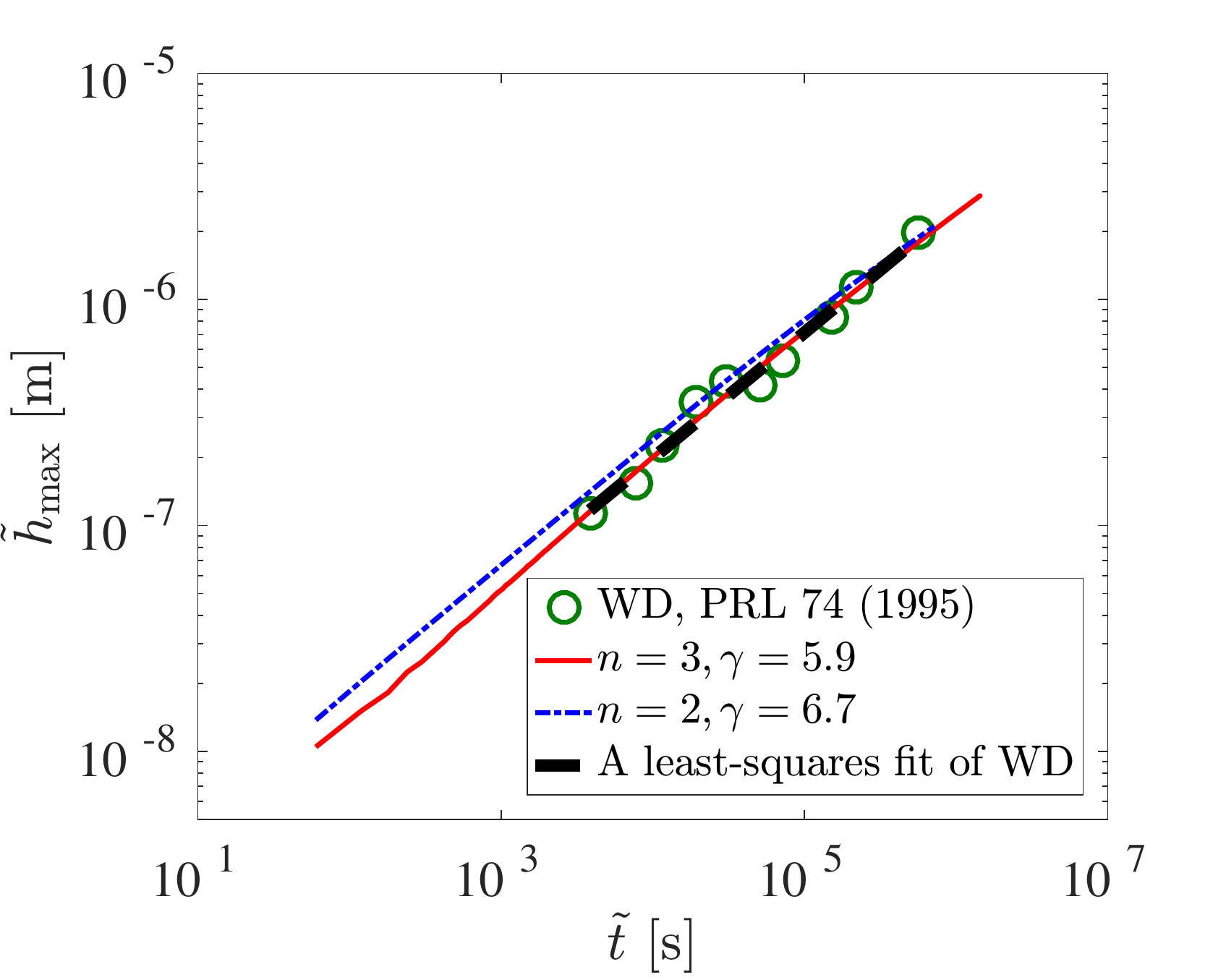} 
\caption{(a) Numerical solutions with the proximity viscosity effects for van der Waals (solid line) and electrostatic (dash-dotted line) interactions are compared to the experimental values described in \cite{Wilen1995} (squares). (b) A least-squares fit of the maximum height of the membrane $\tilde{h}_{\rm max}(\tilde{t})$ obtained from experiments (circles) are compared with that obtained from the numerical solutions for  $d_0 = 3$ \AA, $\mathcal{G} = 0.92 \times 10^2$ K/m, $\eta_r = 10^7$: (i) $n = 3$, $\lambda_n = 1.3759$ \AA, $\gamma = 5.9$, and (ii) $n = 2$, $\lambda_n =0.2101$ \AA, $\gamma = 6.7$.
} 
\label{fig:experiment}
\end{figure}

Finally, we compare the numerical solutions of our model to experiments \cite{Wilen1995}. We choose $d_0 = 3$ \AA ~(the molecular diameter of water is $2.75$ \AA) and experiments on nanometric water films confined between different surfaces show that $\eta_r = 10^7$ \cite{Dhinojwala1997, Major2006}. Thus, using these parameters and the experimental temperature gradient and membrane tension \cite{Wilen1995}, we perform numerical simulations with different $\gamma$ and both van der Waals ($n = 3$) and electrostatic interactions ($n = 2$). For van der Waals interactions, the agreement between theory and experiment is best with $\gamma = 5.9$, whereas for the electrostatic interactions the agreement is best with $\gamma = 6.7$, capturing nearly the entire profile with an overshoot to the maximum deformation for the case of electrostatic interactions (Figure \ref{fig:experiment}). Overall, whereas one might view there to be a better agreement with the data relative to the bulk viscosity model (see figure 4 of \cite{Wettlaufer1996}), here we have an additional parameter, $\gamma$. Next, we compare the numerical solutions with the measurements of the maximum membrane deformation. As shown in Figure \ref{fig:experiment}(b), a least squares fit of the experimental data \cite{Wilen1995} agrees well with our numerical solution with a superior agreement for van der Waals interactions (note again the overshoot in the case of electrostatic interactions), although both are within experimental error. Importantly, we have provided a framework to experimentally test the proximity effect in other materials and to examine the power law treatment of the proximity effect embodied in the exponent $\gamma$. 

Our results have implications for frost heave in the freezing of soils, cells and tissues, cancer treatment, food science, fish biology and botany  \cite{Rubinsky2003, Rempel2007, Peterson2008}, amongst other phase change settings. In particular, our results show that deformation continues even in very low temperatures regions or if the liquid domain is highly restricted, such as in the case of confined bodies subject to temperature gradients, which may rupture (e.g., see \cite{Vlahou2010} and references therein). Moreover, if the temperature gradient is very shallow and the liquid film is thin, deformation can still be large, and our framework provides constraints for designing optimal biological cryogenic preservation \cite{Rubinsky2003}. Finally, among many other biological applications, the processes discussed here are closely related to phase changes and the ordering of water in membranes, which are active areas of inquiry \cite{Parsegian2005}.

\subsection{Generalizations and Implications\label{sec:gen}} 
\textcolor{black}{We have shown here that for $\gamma_{crit} \leqslant \gamma \leqslant \gamma_{sat}$ the mobility term in the flux operator is not scale invariant.  In consequence, because the mobility term depends on the position the similarity solution structure breaks down.  This influence of the proximity effect has implication for a wide class of higher order diffusion-like equations arising in thin film flows, which we can write as
\begin{equation}
\label{eq:general_form}
H_t + \left[ \frac{1}{\eta(H)} F \left(H, H_x, H_{xx}, H_{xxx} \right) \right]_x = 0, 
\end{equation}
where $F$ is a real-valued, nonlinear function of the film thickness $H(x, t)$ and its spatial derivatives up to third oder.   Clearly, the functional form of $F$ and the film-thickness dependent effective viscosity $\eta(H)$ determine the similarity solution structure, as we have shown here as a function of $\gamma$.
We write the confinement effect, as given in equation \eqref{eq:powerlaw_visco_d}, as
$\displaystyle \eta(H) = \eta_b + (\eta_0 - \eta_b) \left( \frac{H_0}{H} \right)^{\gamma}$, where $H_0$ is a structural length scale of the fluid.  Therefore, 
\begin{enumerate}
\item For $\eta(H) \approx \eta_b$, there exists an outer solution to equation \eqref{eq:general_form} away from the region $H \approx H_0$ whenever 
\begin{equation}
(\eta_0 - \eta_b) \left( \frac{H_0}{H} \right)^{\gamma} \ll \eta_b, \qquad \mbox{i.e.,} \qquad \frac{H}{H_0} \gg (\eta_r -1 )^{1/\gamma} > 1. 
\end{equation}
\item For $ \displaystyle \eta(H) \approx (\eta_0 - \eta_b) \left( \frac{H_0}{H} \right)^{\gamma}$, there exists an inner solution to equation \eqref{eq:general_form} near the region $H \approx H_0$ whenever 
\begin{equation}
(\eta_0 - \eta_b) \left( \frac{H_0}{H} \right)^{\gamma} \gg \eta_b, \qquad \mbox{i.e.,} \qquad 1 < \frac{H}{H_0} \ll (\eta_r -1 )^{1/\gamma}. 
\end{equation}
\item The transition from the outer solution to the inner solution occurs on a length scale satisfying 
\begin{equation}
\frac{H}{H_0} \sim (\eta_r -1 )^{1/\gamma} \geqslant 1 \qquad \mbox{for} \qquad \eta_r \geqslant 2, \quad \forall \gamma > 0. 
\end{equation}
\end{enumerate}
This length scale is universal for any thin film problem embodied by \eqref{eq:powerlaw_visco_d} under the influence of this proximity effect.  Thus, this introduces new length scales to problems such as for example droplet breakup, contact line dynamics and wetting and dewetting wherein the finite-time singularity structure of the flux operator will be regularized at larger than traditional (e.g., van der Waals) length scales \cite{EggersPRFluids2018}, as has been seen in recent experiments on moving contact lines in complex fluids \cite{ContactLine}.    
}

\textcolor{black}{A few recent examples of relevance include the inclusion of interface effects in a continuum-based study of transport phenomena in a microfluidic channel by Vo and Kim \cite{VoSciRep2016}, a classical unsteady Stokes flow analysis of oscillating sphere near a wall by Fouxon and Leshansky \cite{FouxonPhysRevE2018}, which would differ at close range where the confinement effect may be operative, and the study of droplet breakup by Deblais \emph{et al.} \cite{DeblaisPhysRevLett2018}.  In this latter combined experimental and numerical study, the authors show inclusion of a small amount of viscosity is sufficient to induce breakdown of the universality of the breakup dynamics predicted by inviscid theory.  Whereas as a general proposition, this is not surprising (namely inviscid theory should in some limit break down), it is a compelling question to examine the relevant length scales of break down and to test the concepts of confinement described here by systematically changing the fluid structure. 
}

\appendix

\section{Low Temperature Boundary Condition \label{sec:BC}}
\begin{figure}[hbtp]
\centering
\includegraphics[scale=0.5]{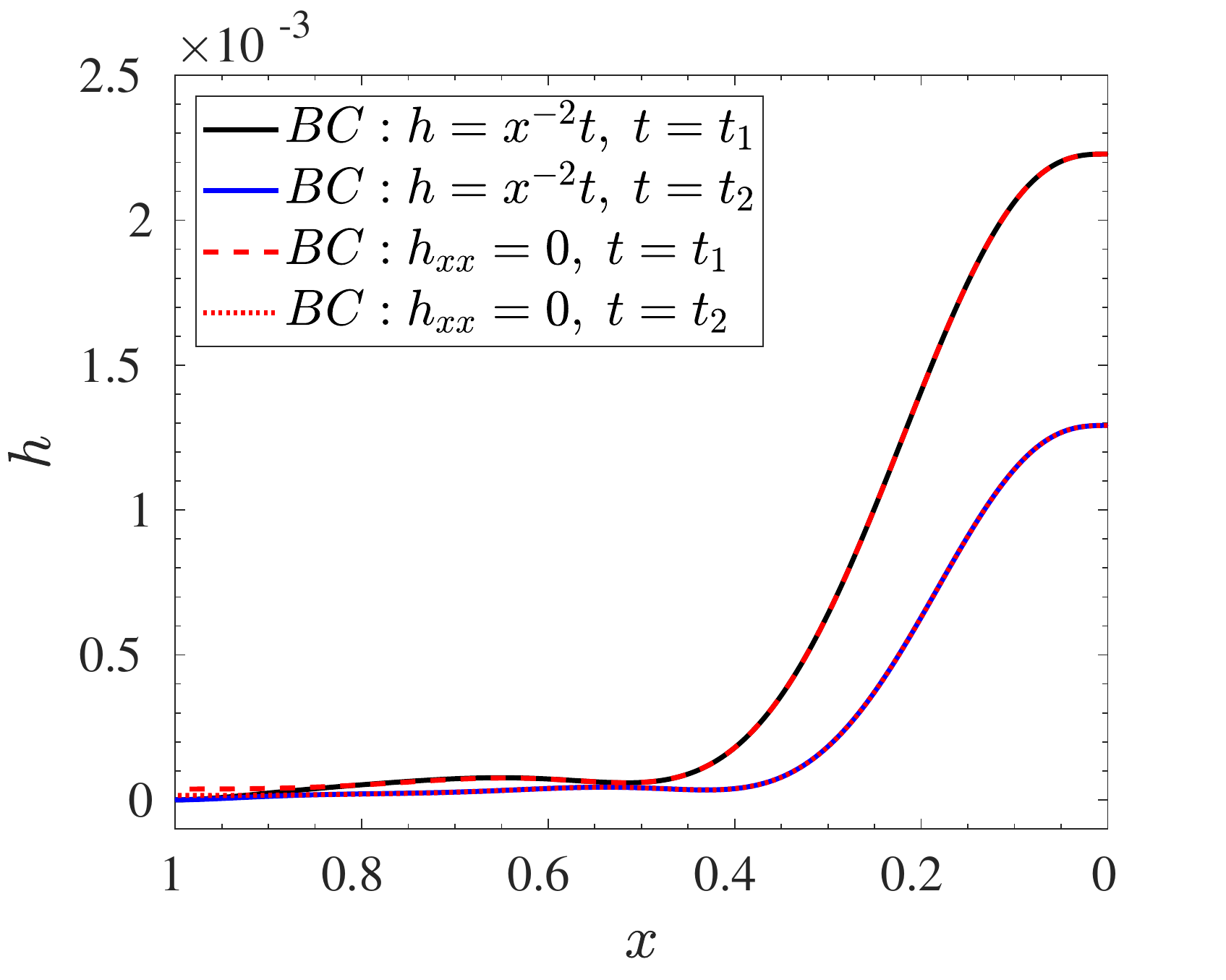} 
\caption{Demonstration of the effects of the low temperature boundary condition for $alpha = 0.6686$ and the bulk viscosity model. Here, $t_1$ and $t_2$ represent two arbitrary dimensionless time.} 
\label{fig:BC_comp}
\end{figure}

As discussed in \S \ref{subsec:nondim} the original theory wherein the bulk viscosity was used the system of ODEs that resulted from the similarity solution given by \eqref{eq:old_similarity_ode}--\eqref{eq:similarity_necessary} was solved \cite{Wettlaufer1996}. Analysis of the evolution equation \eqref{eq:old_similarity_ode} at low temperatures far from the origin lead to the fourth boundary condition, \eqref{eq:similarity_decay}, which is satisfied by a shooting method when solving the system numerically. Here we imposed \eqref{eq:RBC2}. In Figure \ref{fig:BC_comp} we show that there is no change in the membrane height near the warm end and minor changes in the membrane height near the cold end. 

\textcolor{black}{
\section{Approximation of $\gamma_{crit}$ and $\gamma_{sat}$\label{sec:gammas}}
Here, we briefly explain the estimates of the transition values of the parameter $\gamma$; $\gamma_{crit}$ and $\gamma_{sat}$. 
The value of $\gamma_{crit}$ is associated with $\epsilon(\gamma_{crit}) \gg 1$, such that $\tilde{\eta}(\tilde{x})/\eta_b \approx 1$ for $0 \leqslant x \leqslant \varepsilon (\ll 1)$ and $\tilde{\eta}(\tilde{x})/\eta_b \sim \mathbf{O} (x^{\gamma/n})$ for $\varepsilon < x \leqslant 1$.  Thus, 
\begin{eqnarray}
& & (\eta_r - 1) \left[ \frac{d_0}{\lambda_n} \left( \frac{\mathcal{G} X_0}{T_m} \right)^{1/n} \varepsilon^{1/n}  \right]^{\gamma_{crit}} \approx 1, \qquad \textrm{so that}\nonumber \\ 
& & \ln(\eta_r - 1) + \gamma_{crit} \left[ \ln \left( \frac{d_0}{\lambda_n} \right) + \frac{1}{n} \ln \left( \frac{\mathcal{G} X_0}{T_m} \right) + \frac{1}{n} \ln \varepsilon \right] \approx \ln 1 = 0, \qquad \textrm{and hence}\nonumber \\ 
& & \gamma_{crit} \approx - \frac{ \ln(\eta_r - 1) }{ \ln \left( \frac{d_0}{\lambda_n} \right) + \frac{1}{n} \left[ \ln \left( \frac{\mathcal{G} X_0}{T_m} \right) + \ln \varepsilon \right]  \nonumber }. 
\end{eqnarray}
The value of $\gamma_{sat}$ is associated with $\epsilon(\gamma_{sat}) \approx 1$, viz., 
\begin{eqnarray}
& & (\eta_r - 1) \left[ \frac{d_0}{\lambda_n} \left( \frac{\mathcal{G} X_0}{T_m} \right)^{1/n}  \right]^{\gamma_{sat}} \approx 1, \qquad \textrm{so that}\nonumber  \\ 
& & \ln(\eta_r - 1) + \gamma_{sat} \left[ \ln \left( \frac{d_0}{\lambda_n} \right) + \frac{1}{n} \ln \left( \frac{\mathcal{G} X_0}{T_m} \right) \right] \approx \ln 1 = 0, \qquad \textrm{and hence} \nonumber \\ 
& & \gamma_{sat} \approx - \frac{ \ln(\eta_r - 1) }{ \ln \left( \frac{d_0}{\lambda_n} \right) + \frac{1}{n} \ln \left( \frac{\mathcal{G} X_0}{T_m} \right) \nonumber}.
\end{eqnarray} 
}

\section{Code validation\label{sec:code_valid}} 
\textcolor{black}{
We use the five-point stencil formulae for the boundary conditions \eqref{eq:LBC}, \eqref{eq:RBC1}, and \eqref{eq:RBC2} and obtain the following algebraic equations: 
\begin{subequations}
\begin{align}
& h_{-2} - 8 h_{-1} + 8 h_{1} - h_{2} = 0, \label{eq:Warm_BC1_diff} \\ 
- & h_{-2} +16 h_{-1} - 30 h_0 + 16 h_{1} - h_{2} = 0, \label{eq:Warm_BC2_diff} \\ 
& h_{N-2} - 8 h_{N-1} + 8 h_{N+1} - h_{N+1} = 0, \label{eq:Cold_BC1_diff} \\ 
- & h_{N-2} + 16 h_{N-1} - 30 h_{N} + 16 h_{N+1} - h_{N+2} = 0. \label{eq:Cold_BC2_diff} 
\end{align}
\end{subequations}
The constraint of avoiding the singularity at $x = 0$ is checked at every time step using the second order centered difference approximation of the necessary condition \eqref{eq:NC}
\begin{equation}
h_2 - 2h_1 + 2h_{-1} - h_{-2} = -2 \Delta x^3 \alpha^{-1}. \label{eq:Warm_NC_diff} 
\end{equation}
}

\begin{figure}[hbtp]
\centering
\includegraphics[scale=0.5]{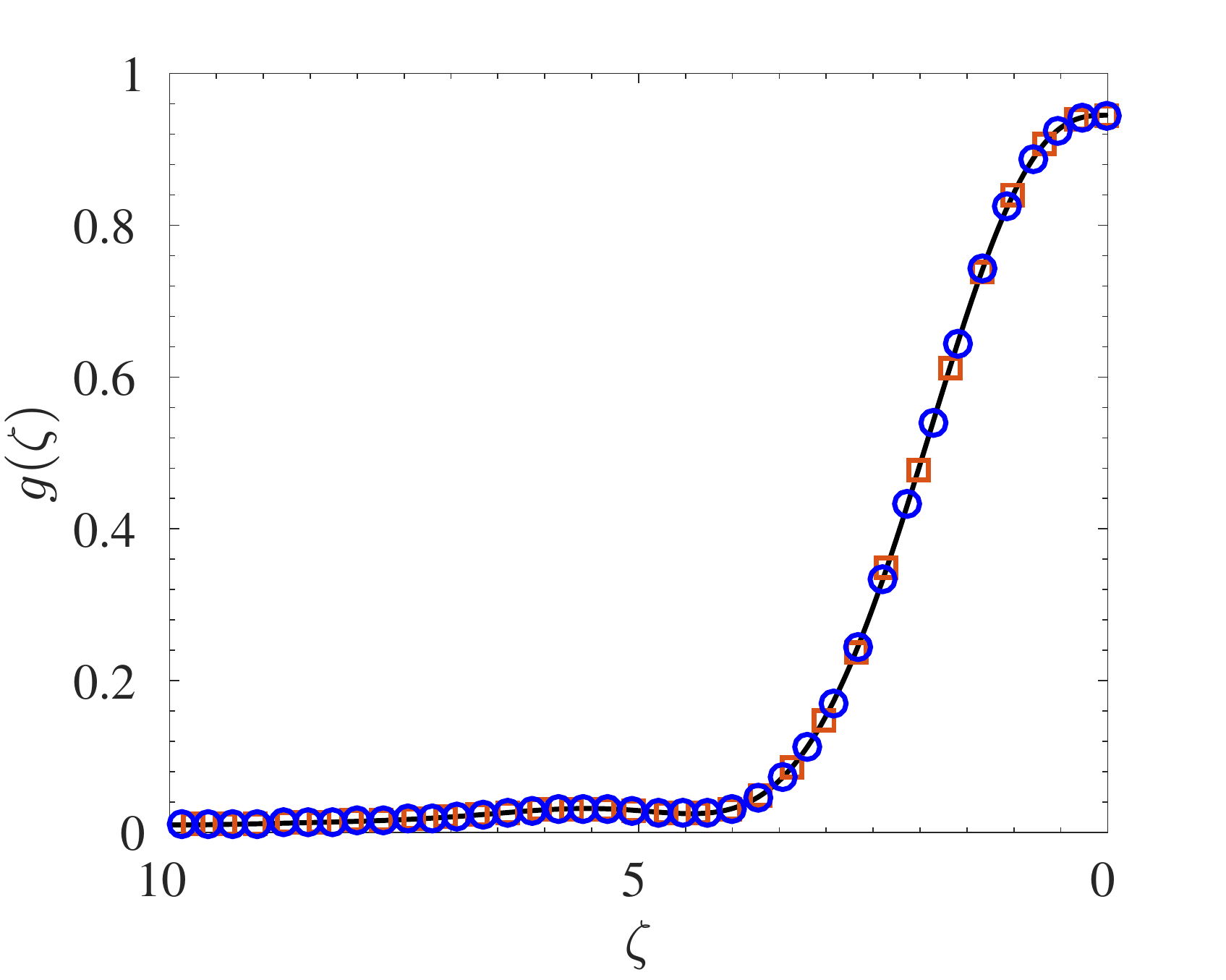} 
\caption{Similarity solutions obtained solving the system \eqref{eq:old_similarity_ode}--\eqref{eq:similarity_necessary} (black solid line) are compared to those obtained from \eqref{eq:mass_conservation_dimless}-\eqref{eq:RBC2} for $\mathcal{G} = 0.92 \times 10^2$ K/m (squares) and $\mathcal{G} = 0.92 \times 10^3$ K/m (circles).} 
\label{fig:similarity1}
\end{figure}

We validate our code by reproducing results from Ref. \cite{Wettlaufer1996}. We solve the partial differential equations corresponding to their constant bulk viscosity model using standard second order finite difference formulae and the time integration is performed using a semi-implicit Crank-Nicolson (C-N) method \cite{LeVeque2007a}. Additionally, we compute the corresponding similarity solution from the numerically computed membrane deformation, $h(x,t)$, using \eqref{eq:old_similarity_sol} and \eqref{eq:old_similarity_var}, as well as solving the corresponding family of ordinary differential equations \eqref{eq:old_similarity_ode}-\eqref{eq:similarity_necessary}. The solutions obtained from these two methods are plotted in Figure \ref{fig:similarity1}, thereby validating our C-N method. 

\begin{figure}[hbtp]
\centering 
(a) \hspace{2.4 in} (b) \\ 
\includegraphics[scale=0.5]{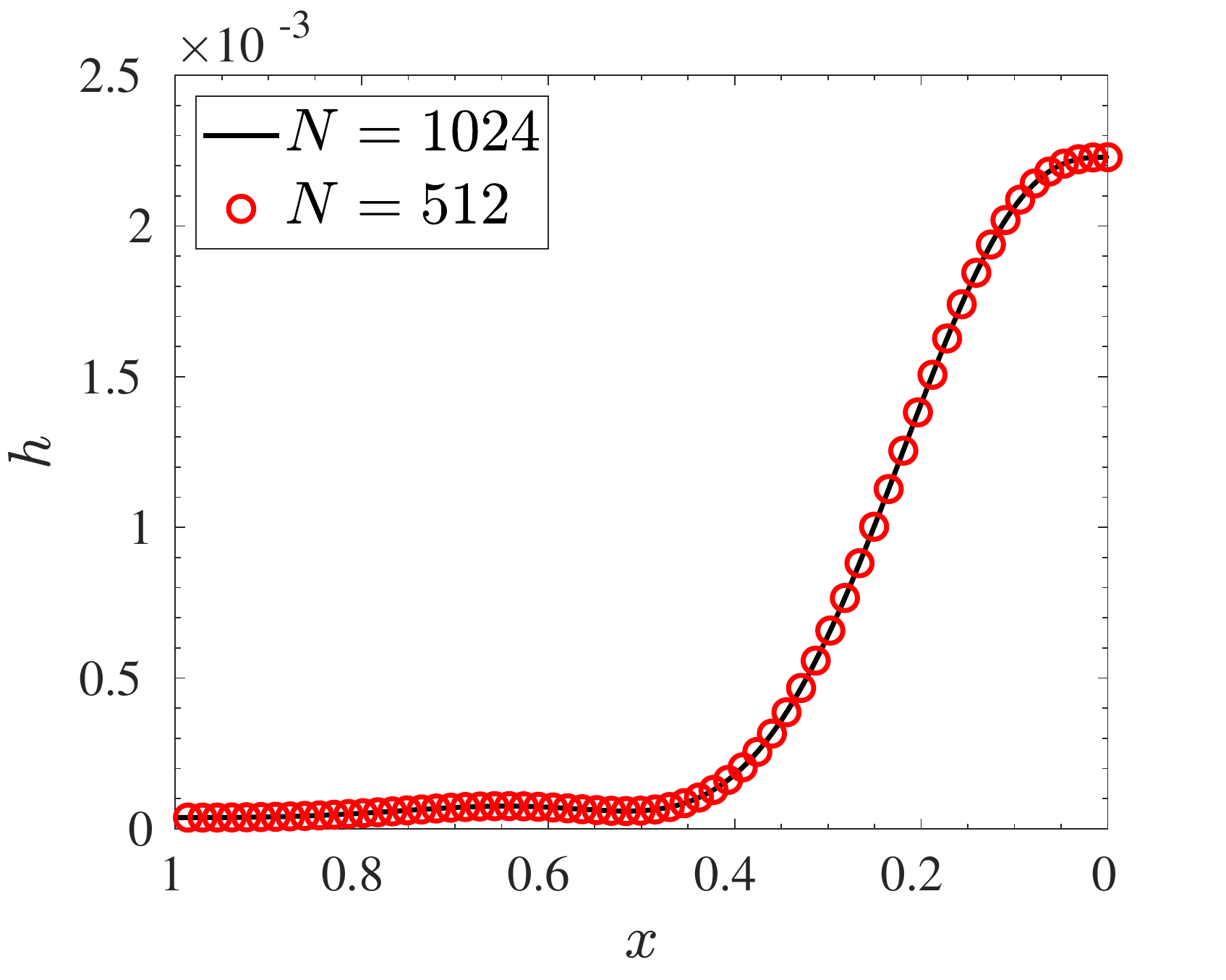} 
\includegraphics[scale=0.5]{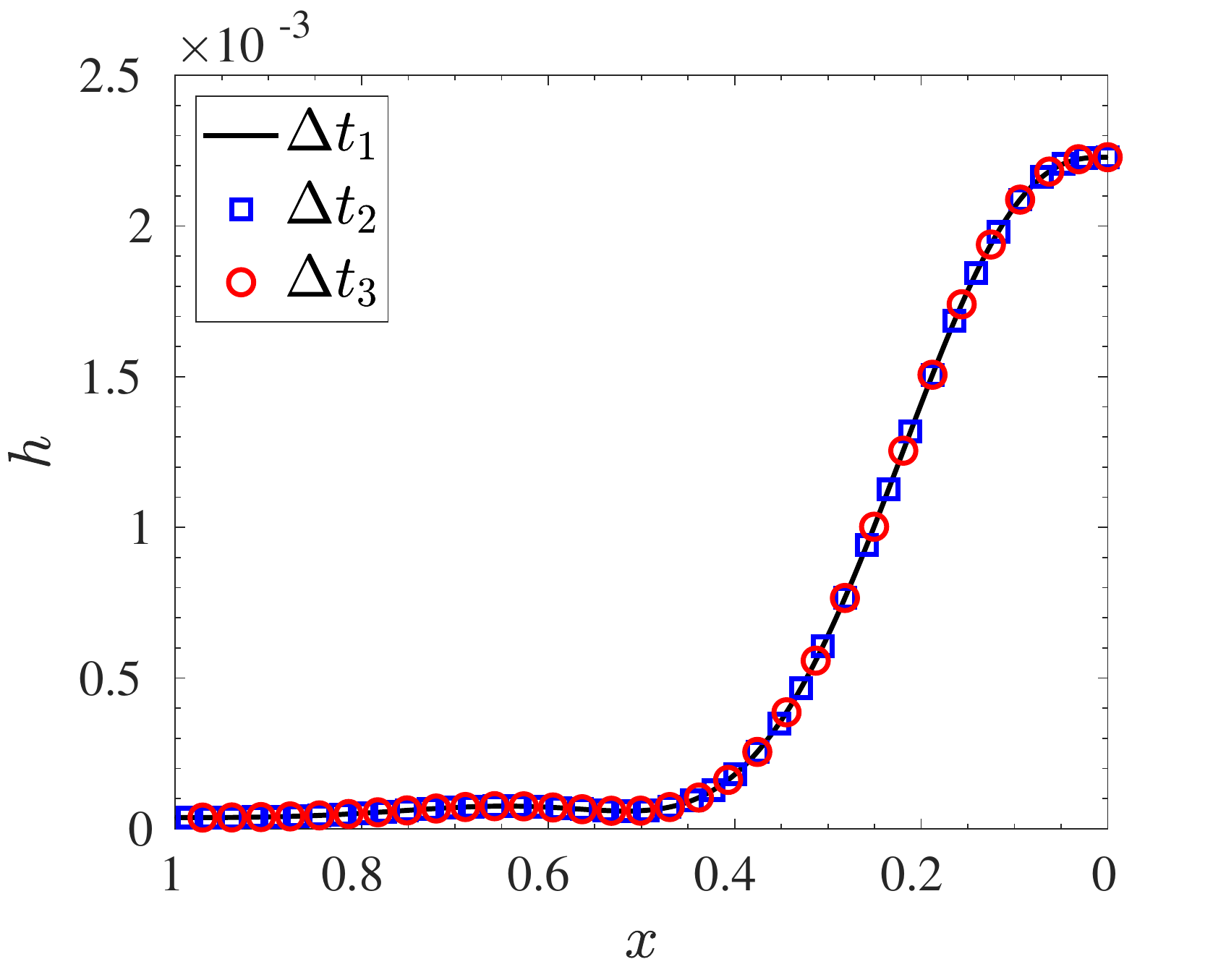} 
\caption{The dimensionless membrane height $h(x,t)$ as a function of the dimensionless position at dimensionless time $t = 3.195 \times 10^{-5}$ for $\alpha = 0.6686$, and the bulk viscosity model. (a) Spatial grid independence for time discretization step $\Delta t_2$. (b) Temporal grid independence for $N = 512$. The time discretization steps $\Delta t_i \; (i = 1, 2, 3)$ satisfy $\Delta t_1/30 = \Delta t_2/10 = \Delta t_3 = 5.5469 \times 10^{-11}$.} 
\label{fig:grid_test}
\end{figure} 

We define the spatial and temporal discretization errors as 
\begin{eqnarray}
\label{eq:x_error}
& & \lVert E_h(t) \rVert = \left( \sum_{i = 1}^{N} \Big\vert h_{\Delta x}(x_i,t) - h_{0.5 \Delta x}(x_{2i},t) \Big\vert^2 \right)^{1/2}, \quad \textrm{and}\\ 
\label{eq:t_error}
& & \lVert E_h(x) \rVert = \left( \sum_{i = 1}^{N} \Big\vert h_{\Delta t_1}(x_i,n_1\Delta t_1) - h_{\Delta t_2}(x_{i},n_2 \Delta t_2) \Big\vert^2 \right)^{1/2}, 
\end{eqnarray}
respectively. Here, $h_{\Delta x}$ represents the numerical solution obtained with a spatial grid size $\Delta x = 1/N$, whereas $h_{0.5 \Delta x}$ is the numerical solution corresponding to the spatial grid size $0.5 \Delta x$, and $n_1, n_2 \in \mathbb{N}$, such that $n_1 \Delta t_1 = n_2 \Delta t_2$. 
Numerical results are spatially (temporally) grid independent if, $\lVert E_h(t) \rVert < \epsilon_1$ uniformly for all $t > 0$ and a specified tolerance $\epsilon_1 > 0$ ($\lVert E_h(x) \rVert < \epsilon_2$, uniformly for all $x \in \Omega$ and a specified tolerance $\epsilon_2 > 0$). For a convergent numerical solution, the optimal spatial and temporal discretization are obtained for $N = 512$ (or $1024$ depending upon various parameters, e.g., $\eta_r, ~ d_0, ~ \lambda_n, ~ \gamma, ~ \mathcal{G}$) and a dimensional value $\Delta t = 10$ s, such that the global discretization error, 
\begin{equation}
\label{eq:global_error}
\epsilon(\Delta x, \Delta t) = \lVert E_h(t) \rVert + \lVert E_h(x) \rVert , 
\end{equation}
is minimized. Figure \ref{fig:grid_test} shows grid independence of $\Delta x$ and $\Delta t$. 

\section{Additional results of reentrant oscillations of $h$ \label{sec:additional}} 
Here we present additional results of parameter dependent reentrant oscillations of the membrane height $\tilde{h}(\tilde{x}, \tilde{t})$. 

\begin{figure}[hbtp]
\centering
(a) \hspace{2.4 in} (b) \\ 
\includegraphics[scale=0.5]{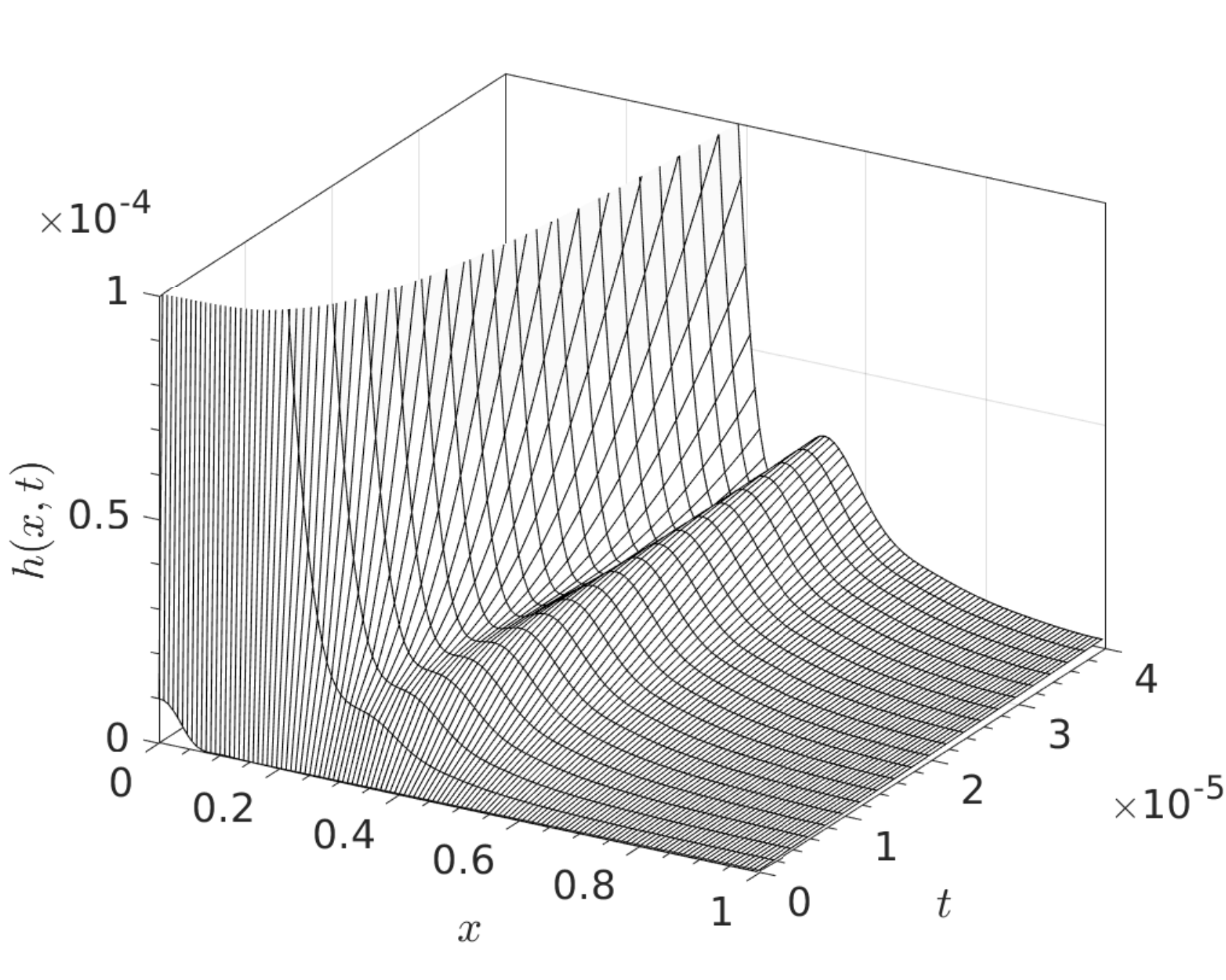}
\includegraphics[scale=0.5]{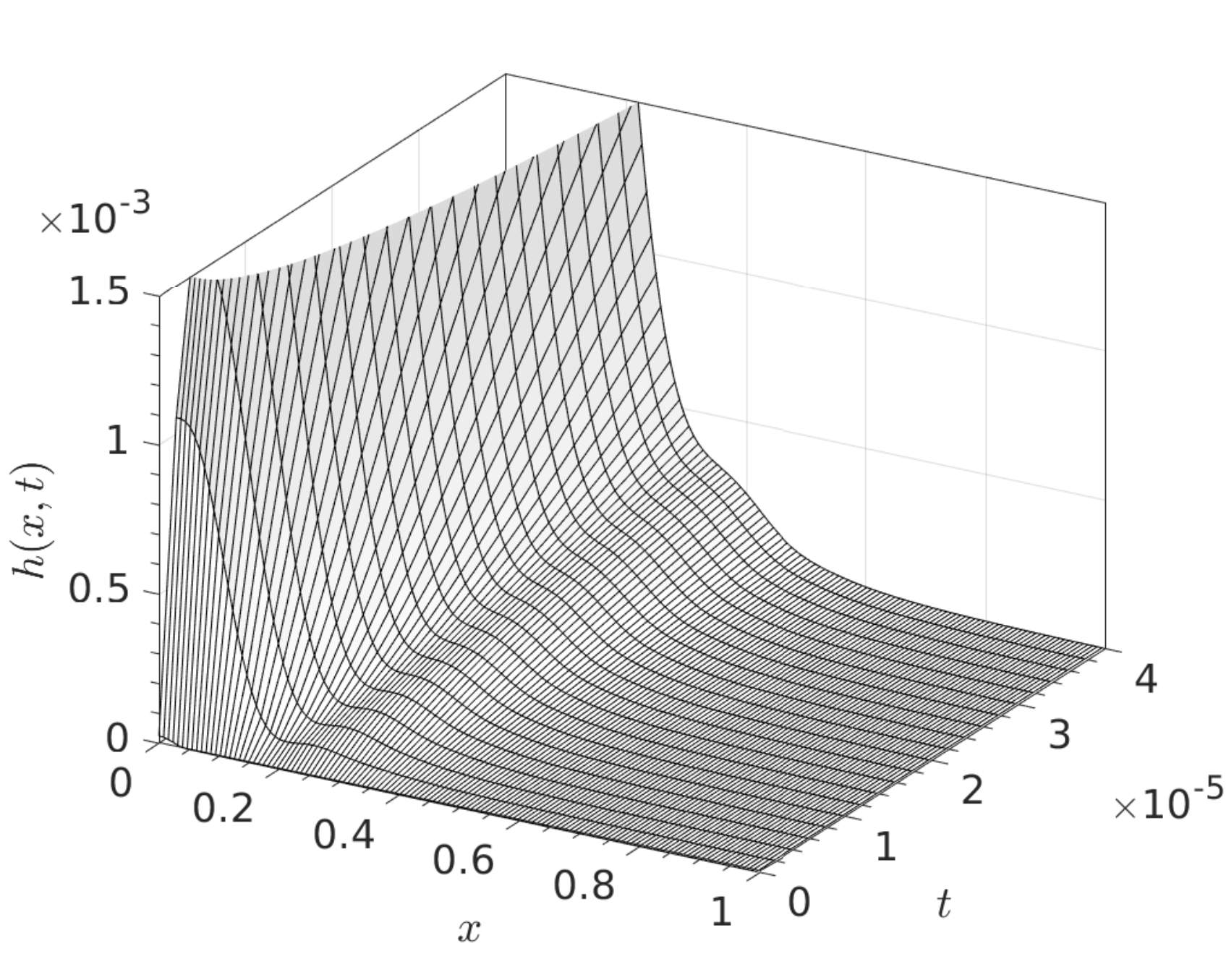}
\caption{(a) Membrane evolution between $h = 0$ and $h = 10^{-4}$, which surround the first minimum and the second maximum, for $\eta_r = 10^7$. $\alpha = 0.6686$, and $d_{8.5, 3} = 8.3973 \times 10^{-6}$. (b) Membrane evolution between $h = 0$ and $h = 1.5 \times 10^{-3}$, which surround the first minimum and the second maximum, for $\eta_r = 10^5$, $\alpha = 0.0669$, and $d_{11.5, 3} = 9.2468 \times 10^{-4}$.} 
\label{fig:surface3}
\end{figure}

\section*{Acknowledgments}
The authors acknowledge the support of the Swedish Research Council Grant No. 638-2013-9243, conversations with their colleagues in Stockholm and New Haven, and J.S.W. acknowledges a Royal Society Wolfson Research Merit Award for support. 


\begin{thebibliography}{58}%
\makeatletter
\providecommand \@ifxundefined [1]{%
 \@ifx{#1\undefined}
}%
\providecommand \@ifnum [1]{%
 \ifnum #1\expandafter \@firstoftwo
 \else \expandafter \@secondoftwo
 \fi
}%
\providecommand \@ifx [1]{%
 \ifx #1\expandafter \@firstoftwo
 \else \expandafter \@secondoftwo
 \fi
}%
\providecommand \natexlab [1]{#1}%
\providecommand \enquote  [1]{``#1''}%
\providecommand \bibnamefont  [1]{#1}%
\providecommand \bibfnamefont [1]{#1}%
\providecommand \citenamefont [1]{#1}%
\providecommand \href@noop [0]{\@secondoftwo}%
\providecommand \href [0]{\begingroup \@sanitize@url \@href}%
\providecommand \@href[1]{\@@startlink{#1}\@@href}%
\providecommand \@@href[1]{\endgroup#1\@@endlink}%
\providecommand \@sanitize@url [0]{\catcode `\\12\catcode `\$12\catcode
  `\&12\catcode `\#12\catcode `\^12\catcode `\_12\catcode `\%12\relax}%
\providecommand \@@startlink[1]{}%
\providecommand \@@endlink[0]{}%
\providecommand \url  [0]{\begingroup\@sanitize@url \@url }%
\providecommand \@url [1]{\endgroup\@href {#1}{\urlprefix }}%
\providecommand \urlprefix  [0]{URL }%
\providecommand \Eprint [0]{\href }%
\providecommand \doibase [0]{http://dx.doi.org/}%
\providecommand \selectlanguage [0]{\@gobble}%
\providecommand \bibinfo  [0]{\@secondoftwo}%
\providecommand \bibfield  [0]{\@secondoftwo}%
\providecommand \translation [1]{[#1]}%
\providecommand \BibitemOpen [0]{}%
\providecommand \bibitemStop [0]{}%
\providecommand \bibitemNoStop [0]{.\EOS\space}%
\providecommand \EOS [0]{\spacefactor3000\relax}%
\providecommand \BibitemShut  [1]{\csname bibitem#1\endcsname}%
\let\auto@bib@innerbib\@empty
\bibitem [{\citenamefont {Wilen}\ \emph {et~al.}(1995)\citenamefont {Wilen},
  \citenamefont {Wettlaufer}, \citenamefont {Elbaum},\ and\ \citenamefont
  {Schick}}]{Wilenetal1995}%
  \BibitemOpen
  \bibfield  {author} {\bibinfo {author} {\bibfnamefont {L.~A.}\ \bibnamefont
  {Wilen}}, \bibinfo {author} {\bibfnamefont {J.~S.}\ \bibnamefont
  {Wettlaufer}}, \bibinfo {author} {\bibfnamefont {M.}~\bibnamefont {Elbaum}},
  \ and\ \bibinfo {author} {\bibfnamefont {M.}~\bibnamefont {Schick}},\
  }\href@noop {} {\bibfield  {journal} {\bibinfo  {journal} {Phys. Rev. B}\
  }\textbf {\bibinfo {volume} {52}},\ \bibinfo {pages} {12,426} (\bibinfo
  {year} {1995})}\BibitemShut {NoStop}%
\bibitem [{\citenamefont {Dash}\ \emph {et~al.}(2006)\citenamefont {Dash},
  \citenamefont {Rempel},\ and\ \citenamefont {Wettlaufer}}]{Dash2006}%
  \BibitemOpen
  \bibfield  {author} {\bibinfo {author} {\bibfnamefont {J.~G.}\ \bibnamefont
  {Dash}}, \bibinfo {author} {\bibfnamefont {A.~W.}\ \bibnamefont {Rempel}}, \
  and\ \bibinfo {author} {\bibfnamefont {J.~S.}\ \bibnamefont {Wettlaufer}},\
  }\href@noop {} {\bibfield  {journal} {\bibinfo  {journal} {Rev. Mod. Phys.}\
  }\textbf {\bibinfo {volume} {78}} (\bibinfo {year} {2006})}\BibitemShut
  {NoStop}%
\bibitem [{\citenamefont {Wettlaufer}\ and\ \citenamefont
  {Worster}(2006)}]{Wettlaufer2006}%
  \BibitemOpen
  \bibfield  {author} {\bibinfo {author} {\bibfnamefont {J.~S.}\ \bibnamefont
  {Wettlaufer}}\ and\ \bibinfo {author} {\bibfnamefont {M.~G.}\ \bibnamefont
  {Worster}},\ }\href@noop {} {\bibfield  {journal} {\bibinfo  {journal} {Annu.
  Rev. Fl. Mech.}\ }\textbf {\bibinfo {volume} {38}},\ \bibinfo {pages} {427}
  (\bibinfo {year} {2006})}\BibitemShut {NoStop}%
\bibitem [{\citenamefont {Schick}(1990)}]{SchickLesHouches}%
  \BibitemOpen
  \bibfield  {author} {\bibinfo {author} {\bibfnamefont {M.}~\bibnamefont
  {Schick}},\ }in\ \href@noop {} {\emph {\bibinfo {booktitle} {Les Houches
  Session XLVIII}}}\ (\bibinfo  {publisher} {Elsevier},\ \bibinfo {address}
  {Amsterdam},\ \bibinfo {year} {1990})\ pp.\ \bibinfo {pages}
  {415--497}\BibitemShut {NoStop}%
\bibitem [{\citenamefont {Safran}(1994)}]{Safranbook}%
  \BibitemOpen
  \bibfield  {author} {\bibinfo {author} {\bibfnamefont {S.~A.}\ \bibnamefont
  {Safran}},\ }\href@noop {} {\emph {\bibinfo {title} {Statistical
  Thermodynamics of Surfaces, Interfaces, and Membranes}}}\ (\bibinfo
  {publisher} {Addison-Wesley},\ \bibinfo {address} {Reading, Massachusetts},\
  \bibinfo {year} {1994})\BibitemShut {NoStop}%
\bibitem [{\citenamefont {French}\ \emph {et~al.}(2010)\citenamefont {French},
  \citenamefont {Parsegian}, \citenamefont {Podgornik}, \citenamefont {Rajter},
  \citenamefont {Jagota}, \citenamefont {Luo}, \citenamefont {Asthagiri},
  \citenamefont {Chaudhury}, \citenamefont {Chiang}, \citenamefont {Granick},
  \citenamefont {Kalinin}, \citenamefont {Kardar}, \citenamefont {Kjellander},
  \citenamefont {Langreth}, \citenamefont {Lewis}, \citenamefont {Lustig},
  \citenamefont {Wesolowski}, \citenamefont {Wettlaufer}, \citenamefont
  {Ching}, \citenamefont {Finnis}, \citenamefont {Houlihan}, \citenamefont {von
  Lilienfeld}, \citenamefont {van Oss},\ and\ \citenamefont
  {Zemb}}]{French:2010}%
  \BibitemOpen
  \bibfield  {author} {\bibinfo {author} {\bibfnamefont {R.~H.}\ \bibnamefont
  {French}}, \bibinfo {author} {\bibfnamefont {V.~A.}\ \bibnamefont
  {Parsegian}}, \bibinfo {author} {\bibfnamefont {R.}~\bibnamefont
  {Podgornik}}, \bibinfo {author} {\bibfnamefont {R.~F.}\ \bibnamefont
  {Rajter}}, \bibinfo {author} {\bibfnamefont {A.}~\bibnamefont {Jagota}},
  \bibinfo {author} {\bibfnamefont {J.}~\bibnamefont {Luo}}, \bibinfo {author}
  {\bibfnamefont {D.}~\bibnamefont {Asthagiri}}, \bibinfo {author}
  {\bibfnamefont {M.~K.}\ \bibnamefont {Chaudhury}}, \bibinfo {author}
  {\bibfnamefont {Y.-m.}\ \bibnamefont {Chiang}}, \bibinfo {author}
  {\bibfnamefont {S.}~\bibnamefont {Granick}}, \bibinfo {author} {\bibfnamefont
  {S.}~\bibnamefont {Kalinin}}, \bibinfo {author} {\bibfnamefont
  {M.}~\bibnamefont {Kardar}}, \bibinfo {author} {\bibfnamefont
  {R.}~\bibnamefont {Kjellander}}, \bibinfo {author} {\bibfnamefont {D.~C.}\
  \bibnamefont {Langreth}}, \bibinfo {author} {\bibfnamefont {J.}~\bibnamefont
  {Lewis}}, \bibinfo {author} {\bibfnamefont {S.}~\bibnamefont {Lustig}},
  \bibinfo {author} {\bibfnamefont {D.}~\bibnamefont {Wesolowski}}, \bibinfo
  {author} {\bibfnamefont {J.~S.}\ \bibnamefont {Wettlaufer}}, \bibinfo
  {author} {\bibfnamefont {W.-Y.}\ \bibnamefont {Ching}}, \bibinfo {author}
  {\bibfnamefont {M.}~\bibnamefont {Finnis}}, \bibinfo {author} {\bibfnamefont
  {F.}~\bibnamefont {Houlihan}}, \bibinfo {author} {\bibfnamefont {O.~A.}\
  \bibnamefont {von Lilienfeld}}, \bibinfo {author} {\bibfnamefont {C.~J.}\
  \bibnamefont {van Oss}}, \ and\ \bibinfo {author} {\bibfnamefont
  {T.}~\bibnamefont {Zemb}},\ }\href {\doibase 10.1103/RevModPhys.82.1887}
  {\bibfield  {journal} {\bibinfo  {journal} {Rev. Mod. Phys.}\ }\textbf
  {\bibinfo {volume} {82}},\ \bibinfo {pages} {1887} (\bibinfo {year}
  {2010})}\BibitemShut {NoStop}%
\bibitem [{\citenamefont {Israelachvili}(2011)}]{Jacobbook}%
  \BibitemOpen
  \bibfield  {author} {\bibinfo {author} {\bibfnamefont {J.~N.}\ \bibnamefont
  {Israelachvili}},\ }\href@noop {} {\emph {\bibinfo {title} {Intermolecular
  and Surface Forces}}},\ \bibinfo {edition} {3rd}\ ed.\ (\bibinfo  {publisher}
  {Academic Press},\ \bibinfo {address} {New York, NY},\ \bibinfo {year}
  {2011})\BibitemShut {NoStop}%
\bibitem [{\citenamefont {Wilen}\ and\ \citenamefont {Dash}(1995)}]{Wilen1995}%
  \BibitemOpen
  \bibfield  {author} {\bibinfo {author} {\bibfnamefont {L.~A.}\ \bibnamefont
  {Wilen}}\ and\ \bibinfo {author} {\bibfnamefont {J.~G.}\ \bibnamefont
  {Dash}},\ }\href {\doibase 10.1103/PhysRevLett.74.5076} {\bibfield  {journal}
  {\bibinfo  {journal} {Phys. Rev. Lett.}\ }\textbf {\bibinfo {volume} {74}},\
  \bibinfo {pages} {5076} (\bibinfo {year} {1995})}\BibitemShut {NoStop}%
\bibitem [{\citenamefont {Wettlaufer}\ and\ \citenamefont
  {Worster}(1995)}]{Wettlaufer1995}%
  \BibitemOpen
  \bibfield  {author} {\bibinfo {author} {\bibfnamefont {J.~S.}\ \bibnamefont
  {Wettlaufer}}\ and\ \bibinfo {author} {\bibfnamefont {M.~G.}\ \bibnamefont
  {Worster}},\ }\href@noop {} {\bibfield  {journal} {\bibinfo  {journal} {Phys.
  Rev. E}\ }\textbf {\bibinfo {volume} {51}},\ \bibinfo {pages} {4679}
  (\bibinfo {year} {1995})}\BibitemShut {NoStop}%
\bibitem [{\citenamefont {Wettlaufer}\ \emph {et~al.}(1996)\citenamefont
  {Wettlaufer}, \citenamefont {Worster}, \citenamefont {Wilen},\ and\
  \citenamefont {Dash}}]{Wettlaufer1996}%
  \BibitemOpen
  \bibfield  {author} {\bibinfo {author} {\bibfnamefont {J.~S.}\ \bibnamefont
  {Wettlaufer}}, \bibinfo {author} {\bibfnamefont {M.~G.}\ \bibnamefont
  {Worster}}, \bibinfo {author} {\bibfnamefont {L.~A.}\ \bibnamefont {Wilen}},
  \ and\ \bibinfo {author} {\bibfnamefont {J.~G.}\ \bibnamefont {Dash}},\
  }\href@noop {} {\bibfield  {journal} {\bibinfo  {journal} {Phys. Rev. Lett.}\
  }\textbf {\bibinfo {volume} {76}},\ \bibinfo {pages} {3602} (\bibinfo {year}
  {1996})}\BibitemShut {NoStop}%
\bibitem [{\citenamefont {Rempel}\ \emph {et~al.}(2004)\citenamefont {Rempel},
  \citenamefont {Wettlaufer},\ and\ \citenamefont {Worster}}]{Rempel2004}%
  \BibitemOpen
  \bibfield  {author} {\bibinfo {author} {\bibfnamefont {A.~W.}\ \bibnamefont
  {Rempel}}, \bibinfo {author} {\bibfnamefont {J.~S.}\ \bibnamefont
  {Wettlaufer}}, \ and\ \bibinfo {author} {\bibfnamefont {M.~G.}\ \bibnamefont
  {Worster}},\ }\href@noop {} {\bibfield  {journal} {\bibinfo  {journal} {J.
  Fluid Mech.}\ }\textbf {\bibinfo {volume} {498}},\ \bibinfo {pages} {227}
  (\bibinfo {year} {2004})}\BibitemShut {NoStop}%
\bibitem [{\citenamefont {Zhu}\ \emph {et~al.}(2000)\citenamefont {Zhu},
  \citenamefont {Vilches}, \citenamefont {Dash}, \citenamefont {Sing},\ and\
  \citenamefont {Wettlaufer}}]{Zhu2000}%
  \BibitemOpen
  \bibfield  {author} {\bibinfo {author} {\bibfnamefont {D.-M.}\ \bibnamefont
  {Zhu}}, \bibinfo {author} {\bibfnamefont {O.~E.}\ \bibnamefont {Vilches}},
  \bibinfo {author} {\bibfnamefont {J.~G.}\ \bibnamefont {Dash}}, \bibinfo
  {author} {\bibfnamefont {B.}~\bibnamefont {Sing}}, \ and\ \bibinfo {author}
  {\bibfnamefont {J.~S.}\ \bibnamefont {Wettlaufer}},\ }\href {\doibase
  10.1103/PhysRevLett.85.4908} {\bibfield  {journal} {\bibinfo  {journal}
  {Phys. Rev. Lett.}\ }\textbf {\bibinfo {volume} {85}},\ \bibinfo {pages}
  {4908} (\bibinfo {year} {2000})}\BibitemShut {NoStop}%
\bibitem [{\citenamefont {Taber}(1929)}]{Taber1929}%
  \BibitemOpen
  \bibfield  {author} {\bibinfo {author} {\bibfnamefont {S.}~\bibnamefont
  {Taber}},\ }\href@noop {} {\bibfield  {journal} {\bibinfo  {journal} {J.
  Geol.}\ }\textbf {\bibinfo {volume} {37}},\ \bibinfo {pages} {428} (\bibinfo
  {year} {1929})}\BibitemShut {NoStop}%
\bibitem [{\citenamefont {Mizusaki}\ and\ \citenamefont
  {Hiroi}(1995)}]{Mizusaki1995}%
  \BibitemOpen
  \bibfield  {author} {\bibinfo {author} {\bibfnamefont {T.}~\bibnamefont
  {Mizusaki}}\ and\ \bibinfo {author} {\bibfnamefont {M.}~\bibnamefont
  {Hiroi}},\ }\href@noop {} {\bibfield  {journal} {\bibinfo  {journal} {Physica
  B}\ }\textbf {\bibinfo {volume} {210}},\ \bibinfo {pages} {403} (\bibinfo
  {year} {1995})}\BibitemShut {NoStop}%
\bibitem [{\citenamefont {Pramanik}\ and\ \citenamefont
  {Wettlaufer}(2017)}]{Pramanik2017}%
  \BibitemOpen
  \bibfield  {author} {\bibinfo {author} {\bibfnamefont {S.}~\bibnamefont
  {Pramanik}}\ and\ \bibinfo {author} {\bibfnamefont {J.~S.}\ \bibnamefont
  {Wettlaufer}},\ }\href {\doibase 10.1103/PhysRevE.96.052801} {\bibfield
  {journal} {\bibinfo  {journal} {Phys. Rev. E}\ }\textbf {\bibinfo {volume}
  {96}},\ \bibinfo {pages} {052801} (\bibinfo {year} {2017})}\BibitemShut
  {NoStop}%
\bibitem [{\citenamefont {Evans}(1990)}]{Evans1990}%
  \BibitemOpen
  \bibfield  {author} {\bibinfo {author} {\bibfnamefont {R.}~\bibnamefont
  {Evans}},\ }\href@noop {} {\bibfield  {journal} {\bibinfo  {journal} {J.
  Phys. Cond. Matt.}\ }\textbf {\bibinfo {volume} {2}},\ \bibinfo {pages}
  {8989} (\bibinfo {year} {1990})}\BibitemShut {NoStop}%
\bibitem [{\citenamefont {Christenson}(2001)}]{Hugo2001}%
  \BibitemOpen
  \bibfield  {author} {\bibinfo {author} {\bibfnamefont {H.~K.}\ \bibnamefont
  {Christenson}},\ }\href@noop {} {\bibfield  {journal} {\bibinfo  {journal}
  {J. Phys. Cond. Matt.}\ }\textbf {\bibinfo {volume} {13}},\ \bibinfo {pages}
  {R95} (\bibinfo {year} {2001})}\BibitemShut {NoStop}%
\bibitem [{\citenamefont {Koplik}\ and\ \citenamefont
  {Banavar}(1995)}]{Koplik1995}%
  \BibitemOpen
  \bibfield  {author} {\bibinfo {author} {\bibfnamefont {J.}~\bibnamefont
  {Koplik}}\ and\ \bibinfo {author} {\bibfnamefont {J.~R.}\ \bibnamefont
  {Banavar}},\ }\href@noop {} {\bibfield  {journal} {\bibinfo  {journal} {Annu.
  Rev. Fl. Mech.}\ }\textbf {\bibinfo {volume} {27}},\ \bibinfo {pages} {257}
  (\bibinfo {year} {1995})}\BibitemShut {NoStop}%
\bibitem [{\citenamefont {Pittenger}\ \emph {et~al.}(2001)\citenamefont
  {Pittenger}, \citenamefont {Fain}, \citenamefont {Cochran}, \citenamefont
  {Donev}, \citenamefont {Robertson}, \citenamefont {Szuchmacher},\ and\
  \citenamefont {Overney}}]{Pittenger2001}%
  \BibitemOpen
  \bibfield  {author} {\bibinfo {author} {\bibfnamefont {B.}~\bibnamefont
  {Pittenger}}, \bibinfo {author} {\bibfnamefont {S.~C.}\ \bibnamefont {Fain}},
  \bibinfo {author} {\bibfnamefont {M.~J.}\ \bibnamefont {Cochran}}, \bibinfo
  {author} {\bibfnamefont {J.~M.~K.}\ \bibnamefont {Donev}}, \bibinfo {author}
  {\bibfnamefont {B.~E.}\ \bibnamefont {Robertson}}, \bibinfo {author}
  {\bibfnamefont {A.}~\bibnamefont {Szuchmacher}}, \ and\ \bibinfo {author}
  {\bibfnamefont {R.~M.}\ \bibnamefont {Overney}},\ }\href@noop {} {\bibfield
  {journal} {\bibinfo  {journal} {Phys. Rev. B}\ }\textbf {\bibinfo {volume}
  {63}},\ \bibinfo {pages} {134102} (\bibinfo {year} {2001})}\BibitemShut
  {NoStop}%
\bibitem [{\citenamefont {D\"{o}ppenschmidt}\ \emph {et~al.}(1998)\citenamefont
  {D\"{o}ppenschmidt}, \citenamefont {Kappl},\ and\ \citenamefont
  {Butt}}]{Butt1998}%
  \BibitemOpen
  \bibfield  {author} {\bibinfo {author} {\bibfnamefont {A.}~\bibnamefont
  {D\"{o}ppenschmidt}}, \bibinfo {author} {\bibfnamefont {M.}~\bibnamefont
  {Kappl}}, \ and\ \bibinfo {author} {\bibfnamefont {H.-J.}\ \bibnamefont
  {Butt}},\ }\href@noop {} {\bibfield  {journal} {\bibinfo  {journal} {J. Phys.
  Chem. B}\ }\textbf {\bibinfo {volume} {102}},\ \bibinfo {pages} {7813}
  (\bibinfo {year} {1998})}\BibitemShut {NoStop}%
\bibitem [{\citenamefont {Dash}(2003)}]{Dash2003}%
  \BibitemOpen
  \bibfield  {author} {\bibinfo {author} {\bibfnamefont {J.~G.}\ \bibnamefont
  {Dash}},\ }\href@noop {} {\bibfield  {journal} {\bibinfo  {journal} {Scripta
  Mat.}\ }\textbf {\bibinfo {volume} {49}},\ \bibinfo {pages} {1003} (\bibinfo
  {year} {2003})}\BibitemShut {NoStop}%
\bibitem [{\citenamefont {Murata}\ \emph {et~al.}(2015)\citenamefont {Murata},
  \citenamefont {Asakawa}, \citenamefont {Nagashima}, \citenamefont
  {Furukawa},\ and\ \citenamefont {Sazaki}}]{Murata2015}%
  \BibitemOpen
  \bibfield  {author} {\bibinfo {author} {\bibfnamefont {K.}~\bibnamefont
  {Murata}}, \bibinfo {author} {\bibfnamefont {H.}~\bibnamefont {Asakawa}},
  \bibinfo {author} {\bibfnamefont {K.}~\bibnamefont {Nagashima}}, \bibinfo
  {author} {\bibfnamefont {Y.}~\bibnamefont {Furukawa}}, \ and\ \bibinfo
  {author} {\bibfnamefont {G.}~\bibnamefont {Sazaki}},\ }\href@noop {}
  {\bibfield  {journal} {\bibinfo  {journal} {Phys. Rev. Lett.}\ }\textbf
  {\bibinfo {volume} {115}},\ \bibinfo {pages} {256103} (\bibinfo {year}
  {2015})}\BibitemShut {NoStop}%
\bibitem [{\citenamefont {Granick}(1991)}]{Granick1991}%
  \BibitemOpen
  \bibfield  {author} {\bibinfo {author} {\bibfnamefont {S.}~\bibnamefont
  {Granick}},\ }\href@noop {} {\bibfield  {journal} {\bibinfo  {journal}
  {Science}\ }\textbf {\bibinfo {volume} {253}},\ \bibinfo {pages} {1374}
  (\bibinfo {year} {1991})}\BibitemShut {NoStop}%
\bibitem [{\citenamefont {Dhinojwala}\ and\ \citenamefont
  {Granick}(1997)}]{Dhinojwala1997}%
  \BibitemOpen
  \bibfield  {author} {\bibinfo {author} {\bibfnamefont {A.}~\bibnamefont
  {Dhinojwala}}\ and\ \bibinfo {author} {\bibfnamefont {S.}~\bibnamefont
  {Granick}},\ }\href {\doibase 10.1021/ja9632318} {\bibfield  {journal}
  {\bibinfo  {journal} {J. Am. Chem. Soc.}\ }\textbf {\bibinfo {volume}
  {119}},\ \bibinfo {pages} {241} (\bibinfo {year} {1997})}\BibitemShut
  {NoStop}%
\bibitem [{\citenamefont {Raviv}\ and\ \citenamefont
  {Klein}(2002)}]{Raviv2002}%
  \BibitemOpen
  \bibfield  {author} {\bibinfo {author} {\bibfnamefont {U.}~\bibnamefont
  {Raviv}}\ and\ \bibinfo {author} {\bibfnamefont {J.}~\bibnamefont {Klein}},\
  }\href@noop {} {\bibfield  {journal} {\bibinfo  {journal} {Science}\ }\textbf
  {\bibinfo {volume} {297}},\ \bibinfo {pages} {1540} (\bibinfo {year}
  {2002})}\BibitemShut {NoStop}%
\bibitem [{\citenamefont {Zhu}\ and\ \citenamefont {Granick}(2003)}]{Zhu2003}%
  \BibitemOpen
  \bibfield  {author} {\bibinfo {author} {\bibfnamefont {Y.}~\bibnamefont
  {Zhu}}\ and\ \bibinfo {author} {\bibfnamefont {S.}~\bibnamefont {Granick}},\
  }\href {\doibase 10.1021/la035155+} {\bibfield  {journal} {\bibinfo
  {journal} {Langmuir}\ }\textbf {\bibinfo {volume} {19}},\ \bibinfo {pages}
  {8148} (\bibinfo {year} {2003})}\BibitemShut {NoStop}%
\bibitem [{\citenamefont {Zhu}\ and\ \citenamefont {Granick}(2004)}]{Zhu2004}%
  \BibitemOpen
  \bibfield  {author} {\bibinfo {author} {\bibfnamefont {Y.}~\bibnamefont
  {Zhu}}\ and\ \bibinfo {author} {\bibfnamefont {S.}~\bibnamefont {Granick}},\
  }\href@noop {} {\bibfield  {journal} {\bibinfo  {journal} {Phys. Rev. Lett.}\
  }\textbf {\bibinfo {volume} {93}},\ \bibinfo {pages} {096101} (\bibinfo
  {year} {2004})}\BibitemShut {NoStop}%
\bibitem [{\citenamefont {Major}\ \emph {et~al.}(2006)\citenamefont {Major},
  \citenamefont {Houston}, \citenamefont {McGrath}, \citenamefont {Siepmann},\
  and\ \citenamefont {Zhu}}]{Major2006}%
  \BibitemOpen
  \bibfield  {author} {\bibinfo {author} {\bibfnamefont {R.~C.}\ \bibnamefont
  {Major}}, \bibinfo {author} {\bibfnamefont {J.~E.}\ \bibnamefont {Houston}},
  \bibinfo {author} {\bibfnamefont {M.~J.}\ \bibnamefont {McGrath}}, \bibinfo
  {author} {\bibfnamefont {J.~I.}\ \bibnamefont {Siepmann}}, \ and\ \bibinfo
  {author} {\bibfnamefont {X.-Y.}\ \bibnamefont {Zhu}},\ }\href@noop {}
  {\bibfield  {journal} {\bibinfo  {journal} {Phys. Rev. Lett.}\ }\textbf
  {\bibinfo {volume} {96}},\ \bibinfo {pages} {177803} (\bibinfo {year}
  {2006})}\BibitemShut {NoStop}%
\bibitem [{\citenamefont {Bureau}(2010)}]{Bureau2010}%
  \BibitemOpen
  \bibfield  {author} {\bibinfo {author} {\bibfnamefont {L.}~\bibnamefont
  {Bureau}},\ }\href {\doibase 10.1103/PhysRevLett.104.218302} {\bibfield
  {journal} {\bibinfo  {journal} {Phys. Rev. Lett.}\ }\textbf {\bibinfo
  {volume} {104}},\ \bibinfo {pages} {218302} (\bibinfo {year}
  {2010})}\BibitemShut {NoStop}%
\bibitem [{\citenamefont {Kessler}\ and\ \citenamefont
  {Werner}(2003)}]{Kessler2003}%
  \BibitemOpen
  \bibfield  {author} {\bibinfo {author} {\bibfnamefont {M.~A.}\ \bibnamefont
  {Kessler}}\ and\ \bibinfo {author} {\bibfnamefont {B.~T.}\ \bibnamefont
  {Werner}},\ }\href {\doibase 10.1126/science.1077309} {\bibfield  {journal}
  {\bibinfo  {journal} {Science}\ }\textbf {\bibinfo {volume} {299}},\ \bibinfo
  {pages} {380} (\bibinfo {year} {2003})}\BibitemShut {NoStop}%
\bibitem [{\citenamefont {Andersland}\ and\ \citenamefont
  {Ladanyi}(2004)}]{Andersland2004}%
  \BibitemOpen
  \bibfield  {author} {\bibinfo {author} {\bibfnamefont {O.~B.}\ \bibnamefont
  {Andersland}}\ and\ \bibinfo {author} {\bibfnamefont {B.}~\bibnamefont
  {Ladanyi}},\ }\href@noop {} {\emph {\bibinfo {title} {An Indroduction to
  Frozen Ground Engineering}}}\ (\bibinfo  {publisher} {Chapman and Hall},\
  \bibinfo {address} {New York, NY},\ \bibinfo {year} {2004})\BibitemShut
  {NoStop}%
\bibitem [{\citenamefont {Rempel}(2007)}]{Rempel2007}%
  \BibitemOpen
  \bibfield  {author} {\bibinfo {author} {\bibfnamefont {A.~W.}\ \bibnamefont
  {Rempel}},\ }\href {\doibase 10.1029/2006JF000525} {\bibfield  {journal}
  {\bibinfo  {journal} {Journal of Geophysical Research: Earth Surface}\
  }\textbf {\bibinfo {volume} {112}},\ \bibinfo {pages} {F02S21} (\bibinfo
  {year} {2007})}\BibitemShut {NoStop}%
\bibitem [{\citenamefont {Peterson}\ and\ \citenamefont
  {Krantz}(2008)}]{Peterson2008}%
  \BibitemOpen
  \bibfield  {author} {\bibinfo {author} {\bibfnamefont {R.~A.}\ \bibnamefont
  {Peterson}}\ and\ \bibinfo {author} {\bibfnamefont {W.~B.}\ \bibnamefont
  {Krantz}},\ }\href@noop {} {\bibfield  {journal} {\bibinfo  {journal}
  {Journal of Geophysical Research: Biogeosciences}\ }\textbf {\bibinfo
  {volume} {113}},\ \bibinfo {pages} {G03S04} (\bibinfo {year}
  {2008})}\BibitemShut {NoStop}%
\bibitem [{\citenamefont {Tamppari}\ \emph {et~al.}(2012)\citenamefont
  {Tamppari}, \citenamefont {Anderson}, \citenamefont {Archer}, \citenamefont
  {Douglas}, \citenamefont {Kounaves}, \citenamefont {McKay}, \citenamefont
  {Ming}, \citenamefont {Moore}, \citenamefont {Quinn}, \citenamefont {Smith},
  \citenamefont {Stroble},\ and\ \citenamefont {Zent}}]{Tamppari:2012}%
  \BibitemOpen
  \bibfield  {author} {\bibinfo {author} {\bibfnamefont {L.~K.}\ \bibnamefont
  {Tamppari}}, \bibinfo {author} {\bibfnamefont {R.~M.}\ \bibnamefont
  {Anderson}}, \bibinfo {author} {\bibfnamefont {J.}~\bibnamefont {Archer},
  \bibfnamefont {P.~D.}}, \bibinfo {author} {\bibfnamefont {S.}~\bibnamefont
  {Douglas}}, \bibinfo {author} {\bibfnamefont {S.~P.}\ \bibnamefont
  {Kounaves}}, \bibinfo {author} {\bibfnamefont {C.~P.}\ \bibnamefont {McKay}},
  \bibinfo {author} {\bibfnamefont {D.~W.}\ \bibnamefont {Ming}}, \bibinfo
  {author} {\bibfnamefont {Q.}~\bibnamefont {Moore}}, \bibinfo {author}
  {\bibfnamefont {J.~E.}\ \bibnamefont {Quinn}}, \bibinfo {author}
  {\bibfnamefont {P.~H.}\ \bibnamefont {Smith}}, \bibinfo {author}
  {\bibfnamefont {S.}~\bibnamefont {Stroble}}, \ and\ \bibinfo {author}
  {\bibfnamefont {A.~P.}\ \bibnamefont {Zent}},\ }\href@noop {} {\bibfield
  {journal} {\bibinfo  {journal} {Antarct. Sci.}\ }\textbf {\bibinfo {volume}
  {24}},\ \bibinfo {pages} {211} (\bibinfo {year} {2012})}\BibitemShut
  {NoStop}%
\bibitem [{\citenamefont {Rubinsky}(2003)}]{Rubinsky2003}%
  \BibitemOpen
  \bibfield  {author} {\bibinfo {author} {\bibfnamefont {B.}~\bibnamefont
  {Rubinsky}},\ }\href {\doibase 10.1023/A:1024734003814} {\bibfield  {journal}
  {\bibinfo  {journal} {Heart Failure Reviews}\ }\textbf {\bibinfo {volume}
  {8}},\ \bibinfo {pages} {277} (\bibinfo {year} {2003})}\BibitemShut {NoStop}%
\bibitem [{\citenamefont {Hawthorn}\ and\ \citenamefont
  {Rolfe}(2016)}]{Hawthorn2016}%
  \BibitemOpen
  \bibfield  {author} {\bibinfo {author} {\bibfnamefont {J.}~\bibnamefont
  {Hawthorn}}\ and\ \bibinfo {author} {\bibfnamefont {E.}~\bibnamefont
  {Rolfe}},\ }\href@noop {} {\emph {\bibinfo {title} {Low Temperature Biology
  of Foodstuffs: Recent Advances in Food Science}}},\ Recent advances in food
  science\ (\bibinfo  {publisher} {Elsevier Science},\ \bibinfo {year}
  {2016})\BibitemShut {NoStop}%
\bibitem [{\citenamefont {Ishiguro}\ and\ \citenamefont
  {Rubinsky}(1994)}]{Ishiguro1994}%
  \BibitemOpen
  \bibfield  {author} {\bibinfo {author} {\bibfnamefont {H.}~\bibnamefont
  {Ishiguro}}\ and\ \bibinfo {author} {\bibfnamefont {B.}~\bibnamefont
  {Rubinsky}},\ }\href@noop {} {\bibfield  {journal} {\bibinfo  {journal}
  {Cryobiology}\ }\textbf {\bibinfo {volume} {31}},\ \bibinfo {pages} {483 }
  (\bibinfo {year} {1994})}\BibitemShut {NoStop}%
\bibitem [{\citenamefont {Bar-Dolev}\ \emph {et~al.}(2012)\citenamefont
  {Bar-Dolev}, \citenamefont {Celik}, \citenamefont {Wettlaufer}, \citenamefont
  {Davies},\ and\ \citenamefont {Braslavsky}}]{Bar-Dolev:2012}%
  \BibitemOpen
  \bibfield  {author} {\bibinfo {author} {\bibfnamefont {M.}~\bibnamefont
  {Bar-Dolev}}, \bibinfo {author} {\bibfnamefont {Y.}~\bibnamefont {Celik}},
  \bibinfo {author} {\bibfnamefont {J.~S.}\ \bibnamefont {Wettlaufer}},
  \bibinfo {author} {\bibfnamefont {P.~L.}\ \bibnamefont {Davies}}, \ and\
  \bibinfo {author} {\bibfnamefont {I.}~\bibnamefont {Braslavsky}},\ }\href
  {\doibase 10.1098/rsif.2012.0388} {\bibfield  {journal} {\bibinfo  {journal}
  {J. R. Soc. Interface}\ }\textbf {\bibinfo {volume} {9}},\ \bibinfo {pages}
  {3249} (\bibinfo {year} {2012})}\BibitemShut {NoStop}%
\bibitem [{\citenamefont {Rubinsky}\ and\ \citenamefont
  {Pegg}(1988)}]{Rubinsky1988}%
  \BibitemOpen
  \bibfield  {author} {\bibinfo {author} {\bibfnamefont {B.}~\bibnamefont
  {Rubinsky}}\ and\ \bibinfo {author} {\bibfnamefont {D.~E.}\ \bibnamefont
  {Pegg}},\ }\href {\doibase 10.1098/rspb.1988.0053} {\bibfield  {journal}
  {\bibinfo  {journal} {Proceedings of the Royal Society of London B:
  Biological Sciences}\ }\textbf {\bibinfo {volume} {234}},\ \bibinfo {pages}
  {343} (\bibinfo {year} {1988})}\BibitemShut {NoStop}%
\bibitem [{\citenamefont {Huppert}(1982)}]{Huppert1982}%
  \BibitemOpen
  \bibfield  {author} {\bibinfo {author} {\bibfnamefont {H.~E.}\ \bibnamefont
  {Huppert}},\ }\href@noop {} {\bibfield  {journal} {\bibinfo  {journal}
  {Nature}\ }\textbf {\bibinfo {volume} {300}},\ \bibinfo {pages} {427}
  (\bibinfo {year} {1982})}\BibitemShut {NoStop}%
\bibitem [{\citenamefont {Troian}\ \emph {et~al.}(1989)\citenamefont {Troian},
  \citenamefont {Herbolzheimer}, \citenamefont {Safran},\ and\ \citenamefont
  {Joanny}}]{Troian1989}%
  \BibitemOpen
  \bibfield  {author} {\bibinfo {author} {\bibfnamefont {S.~M.}\ \bibnamefont
  {Troian}}, \bibinfo {author} {\bibfnamefont {E.}~\bibnamefont
  {Herbolzheimer}}, \bibinfo {author} {\bibfnamefont {S.~A.}\ \bibnamefont
  {Safran}}, \ and\ \bibinfo {author} {\bibfnamefont {J.~F.}\ \bibnamefont
  {Joanny}},\ }\href@noop {} {\bibfield  {journal} {\bibinfo  {journal} {EPL
  (Europhysics Letters)}\ }\textbf {\bibinfo {volume} {10}},\ \bibinfo {pages}
  {25} (\bibinfo {year} {1989})}\BibitemShut {NoStop}%
\bibitem [{\citenamefont {Oron}\ \emph {et~al.}(1997)\citenamefont {Oron},
  \citenamefont {Davis},\ and\ \citenamefont {Bankoff}}]{Oron:1997}%
  \BibitemOpen
  \bibfield  {author} {\bibinfo {author} {\bibfnamefont {A.}~\bibnamefont
  {Oron}}, \bibinfo {author} {\bibfnamefont {S.~H.}\ \bibnamefont {Davis}}, \
  and\ \bibinfo {author} {\bibfnamefont {S.~G.}\ \bibnamefont {Bankoff}},\
  }\href@noop {} {\bibfield  {journal} {\bibinfo  {journal} {Rev. Mod. Phys.}\
  }\textbf {\bibinfo {volume} {69}},\ \bibinfo {pages} {931} (\bibinfo {year}
  {1997})}\BibitemShut {NoStop}%
\bibitem [{\citenamefont {Craster}\ and\ \citenamefont
  {Matar}(2009)}]{Craster:2009}%
  \BibitemOpen
  \bibfield  {author} {\bibinfo {author} {\bibfnamefont {R.~V.}\ \bibnamefont
  {Craster}}\ and\ \bibinfo {author} {\bibfnamefont {O.~K.}\ \bibnamefont
  {Matar}},\ }\href {\doibase 10.1103/RevModPhys.81.1131} {\bibfield  {journal}
  {\bibinfo  {journal} {Rev. Mod. Phys.}\ }\textbf {\bibinfo {volume} {81}},\
  \bibinfo {pages} {1131} (\bibinfo {year} {2009})}\BibitemShut {NoStop}%
\bibitem [{\citenamefont {Constantin}\ \emph {et~al.}(1993)\citenamefont
  {Constantin}, \citenamefont {Dupont}, \citenamefont {Goldstein},
  \citenamefont {Kadanoff}, \citenamefont {Shelley},\ and\ \citenamefont
  {Zhou}}]{Constantin:1993}%
  \BibitemOpen
  \bibfield  {author} {\bibinfo {author} {\bibfnamefont {P.}~\bibnamefont
  {Constantin}}, \bibinfo {author} {\bibfnamefont {T.~F.}\ \bibnamefont
  {Dupont}}, \bibinfo {author} {\bibfnamefont {R.~E.}\ \bibnamefont
  {Goldstein}}, \bibinfo {author} {\bibfnamefont {L.~P.}\ \bibnamefont
  {Kadanoff}}, \bibinfo {author} {\bibfnamefont {M.~J.}\ \bibnamefont
  {Shelley}}, \ and\ \bibinfo {author} {\bibfnamefont {S.~M.}\ \bibnamefont
  {Zhou}},\ }\href@noop {} {\bibfield  {journal} {\bibinfo  {journal} {Phys.
  Rev. E}\ }\textbf {\bibinfo {volume} {47}},\ \bibinfo {pages} {4169}
  (\bibinfo {year} {1993})}\BibitemShut {NoStop}%
\bibitem [{\citenamefont {Hewitt}\ \emph {et~al.}(2015)\citenamefont {Hewitt},
  \citenamefont {Balmforth},\ and\ \citenamefont {De~Bruyn}}]{Hewitt2015}%
  \BibitemOpen
  \bibfield  {author} {\bibinfo {author} {\bibfnamefont {I.~J.}\ \bibnamefont
  {Hewitt}}, \bibinfo {author} {\bibfnamefont {N.~J.}\ \bibnamefont
  {Balmforth}}, \ and\ \bibinfo {author} {\bibfnamefont {J.~R.}\ \bibnamefont
  {De~Bruyn}},\ }\href@noop {} {\bibfield  {journal} {\bibinfo  {journal}
  {European Journal of Applied Mathematics}\ }\textbf {\bibinfo {volume}
  {26}},\ \bibinfo {pages} {1} (\bibinfo {year} {2015})}\BibitemShut {NoStop}%
\bibitem [{\citenamefont {Bertozzi}\ and\ \citenamefont
  {Pugh}(2000)}]{Bertozzi2000}%
  \BibitemOpen
  \bibfield  {author} {\bibinfo {author} {\bibfnamefont {A.~L.}\ \bibnamefont
  {Bertozzi}}\ and\ \bibinfo {author} {\bibfnamefont {M.~C.}\ \bibnamefont
  {Pugh}},\ }\href@noop {} {\bibfield  {journal} {\bibinfo  {journal} {Indiana
  University Mathematics Journal}\ }\textbf {\bibinfo {volume} {49}},\ \bibinfo
  {pages} {1323} (\bibinfo {year} {2000})}\BibitemShut {NoStop}%
\bibitem [{\citenamefont {Nada}\ and\ \citenamefont
  {Furukawa}(1997)}]{Nada:1997}%
  \BibitemOpen
  \bibfield  {author} {\bibinfo {author} {\bibfnamefont {H.}~\bibnamefont
  {Nada}}\ and\ \bibinfo {author} {\bibfnamefont {Y.}~\bibnamefont
  {Furukawa}},\ }\href@noop {} {\bibfield  {journal} {\bibinfo  {journal} {J.
  Phys. Chem. B}\ }\textbf {\bibinfo {volume} {101}},\ \bibinfo {pages} {6163}
  (\bibinfo {year} {1997})}\BibitemShut {NoStop}%
\bibitem [{\citenamefont {Nada}\ and\ \citenamefont
  {Furukawa}(2000)}]{Nada:2000}%
  \BibitemOpen
  \bibfield  {author} {\bibinfo {author} {\bibfnamefont {H.}~\bibnamefont
  {Nada}}\ and\ \bibinfo {author} {\bibfnamefont {Y.}~\bibnamefont
  {Furukawa}},\ }\href@noop {} {\bibfield  {journal} {\bibinfo  {journal}
  {Surf. Sci.}\ }\textbf {\bibinfo {volume} {446}},\ \bibinfo {pages} {1}
  (\bibinfo {year} {2000})}\BibitemShut {NoStop}%
\bibitem [{\citenamefont {Nada}(2016)}]{Nada:2016}%
  \BibitemOpen
  \bibfield  {author} {\bibinfo {author} {\bibfnamefont {H.}~\bibnamefont
  {Nada}},\ }\href {\doibase 10.1063/1.4973000} {\bibfield  {journal} {\bibinfo
   {journal} {J. Chem. Phys.}\ }\textbf {\bibinfo {volume} {145}} (\bibinfo
  {year} {2016}),\ 10.1063/1.4973000}\BibitemShut {NoStop}%
\bibitem [{\citenamefont {LeVeque}(2007)}]{LeVeque2007a}%
  \BibitemOpen
  \bibfield  {author} {\bibinfo {author} {\bibfnamefont {R.}~\bibnamefont
  {LeVeque}},\ }\href@noop {} {\emph {\bibinfo {title} {Finite Difference
  Methods for Ordinary and Partial Differential Equations}}}\ (\bibinfo
  {publisher} {Society for Industrial and Applied Mathematics},\ \bibinfo
  {year} {2007})\BibitemShut {NoStop}%
\bibitem [{\citenamefont {Bertozzi}(1998)}]{Bertozzi:1998}%
  \BibitemOpen
  \bibfield  {author} {\bibinfo {author} {\bibfnamefont {A.~L.}\ \bibnamefont
  {Bertozzi}},\ }\href@noop {} {\bibfield  {journal} {\bibinfo  {journal} {Not.
  Am. Math. Soc.}\ }\textbf {\bibinfo {volume} {45}},\ \bibinfo {pages} {689}
  (\bibinfo {year} {1998})}\BibitemShut {NoStop}%
\bibitem [{\citenamefont {Chapman}\ \emph {et~al.}(2013)\citenamefont
  {Chapman}, \citenamefont {Trinh},\ and\ \citenamefont
  {Witelski}}]{Chapman:2013fk}%
  \BibitemOpen
  \bibfield  {author} {\bibinfo {author} {\bibfnamefont {S.~J.}\ \bibnamefont
  {Chapman}}, \bibinfo {author} {\bibfnamefont {P.~H.}\ \bibnamefont {Trinh}},
  \ and\ \bibinfo {author} {\bibfnamefont {T.~P.}\ \bibnamefont {Witelski}},\
  }\href {\doibase 10.1137/120872012} {\bibfield  {journal} {\bibinfo
  {journal} {SIAM J. Appl. Math.}\ }\textbf {\bibinfo {volume} {73}},\ \bibinfo
  {pages} {232} (\bibinfo {year} {2013})}\BibitemShut {NoStop}%
\bibitem [{\citenamefont {Vlahou}\ and\ \citenamefont
  {Worster}(2010)}]{Vlahou2010}%
  \BibitemOpen
  \bibfield  {author} {\bibinfo {author} {\bibfnamefont {I.}~\bibnamefont
  {Vlahou}}\ and\ \bibinfo {author} {\bibfnamefont {M.~G.}\ \bibnamefont
  {Worster}},\ }\href@noop {} {\bibfield  {journal} {\bibinfo  {journal} {J.
  Glaciol.}\ }\textbf {\bibinfo {volume} {56}},\ \bibinfo {pages} {271}
  (\bibinfo {year} {2010})}\BibitemShut {NoStop}%
\bibitem [{\citenamefont {Parsegian}(2005)}]{Parsegian2005}%
  \BibitemOpen
  \bibfield  {author} {\bibinfo {author} {\bibfnamefont {V.~A.}\ \bibnamefont
  {Parsegian}},\ }\href@noop {} {\emph {\bibinfo {title} {Van der Waals Forces:
  A Handbook for Biologists, Chemists, Engineers, and Physicists}}}\ (\bibinfo
  {publisher} {Cambridge University Press},\ \bibinfo {year}
  {2005})\BibitemShut {NoStop}%
\bibitem [{\citenamefont {Eggers}(2018)}]{EggersPRFluids2018}%
  \BibitemOpen
  \bibfield  {author} {\bibinfo {author} {\bibfnamefont {J.}~\bibnamefont
  {Eggers}},\ }\href {\doibase 10.1103/PhysRevFluids.3.110503} {\bibfield
  {journal} {\bibinfo  {journal} {Phys. Rev. Fluids}\ }\textbf {\bibinfo
  {volume} {3}},\ \bibinfo {pages} {110503} (\bibinfo {year}
  {2018})}\BibitemShut {NoStop}%
\bibitem [{\citenamefont {Nanjundiah}\ \emph {et~al.}(2019)\citenamefont
  {Nanjundiah}, \citenamefont {Kurian}, \citenamefont {Kaur}, \citenamefont
  {Singla},\ and\ \citenamefont {Dhinojwala}}]{ContactLine}%
  \BibitemOpen
  \bibfield  {author} {\bibinfo {author} {\bibfnamefont {K.}~\bibnamefont
  {Nanjundiah}}, \bibinfo {author} {\bibfnamefont {A.}~\bibnamefont {Kurian}},
  \bibinfo {author} {\bibfnamefont {S.}~\bibnamefont {Kaur}}, \bibinfo {author}
  {\bibfnamefont {S.}~\bibnamefont {Singla}}, \ and\ \bibinfo {author}
  {\bibfnamefont {A.}~\bibnamefont {Dhinojwala}},\ }\href@noop {} {\bibfield
  {journal} {\bibinfo  {journal} {Phys. Rev. Lett.}\ }\textbf {\bibinfo
  {volume} {122}} (\bibinfo {year} {2019})}\BibitemShut {NoStop}%
\bibitem [{\citenamefont {Vo}\ and\ \citenamefont {Kim}(2016)}]{VoSciRep2016}%
  \BibitemOpen
  \bibfield  {author} {\bibinfo {author} {\bibfnamefont {T.~Q.}\ \bibnamefont
  {Vo}}\ and\ \bibinfo {author} {\bibfnamefont {B.}~\bibnamefont {Kim}},\
  }\href {\doibase 10.1038/srep33881} {\bibfield  {journal} {\bibinfo
  {journal} {Sci. Rep.}\ }\textbf {\bibinfo {volume} {6}},\ \bibinfo {pages}
  {33881} (\bibinfo {year} {2016})}\BibitemShut {NoStop}%
\bibitem [{\citenamefont {Fouxon}\ and\ \citenamefont
  {Leshansky}(2018)}]{FouxonPhysRevE2018}%
  \BibitemOpen
  \bibfield  {author} {\bibinfo {author} {\bibfnamefont {I.}~\bibnamefont
  {Fouxon}}\ and\ \bibinfo {author} {\bibfnamefont {A.}~\bibnamefont
  {Leshansky}},\ }\href {\doibase 10.1103/PhysRevE.98.063108} {\bibfield
  {journal} {\bibinfo  {journal} {Phys. Rev. E}\ }\textbf {\bibinfo {volume}
  {98}},\ \bibinfo {pages} {063108} (\bibinfo {year} {2018})}\BibitemShut
  {NoStop}%
\bibitem [{\citenamefont {Deblais}\ \emph {et~al.}(2018)\citenamefont
  {Deblais}, \citenamefont {Herrada}, \citenamefont {Hauner}, \citenamefont
  {Velikov}, \citenamefont {van Roon}, \citenamefont {Kellay}, \citenamefont
  {Eggers},\ and\ \citenamefont {Bonn}}]{DeblaisPhysRevLett2018}%
  \BibitemOpen
  \bibfield  {author} {\bibinfo {author} {\bibfnamefont {A.}~\bibnamefont
  {Deblais}}, \bibinfo {author} {\bibfnamefont {M.~A.}\ \bibnamefont
  {Herrada}}, \bibinfo {author} {\bibfnamefont {I.}~\bibnamefont {Hauner}},
  \bibinfo {author} {\bibfnamefont {K.~P.}\ \bibnamefont {Velikov}}, \bibinfo
  {author} {\bibfnamefont {T.}~\bibnamefont {van Roon}}, \bibinfo {author}
  {\bibfnamefont {H.}~\bibnamefont {Kellay}}, \bibinfo {author} {\bibfnamefont
  {J.}~\bibnamefont {Eggers}}, \ and\ \bibinfo {author} {\bibfnamefont
  {D.}~\bibnamefont {Bonn}},\ }\href {\doibase 10.1103/PhysRevLett.121.254501}
  {\bibfield  {journal} {\bibinfo  {journal} {Phys. Rev. Lett.}\ }\textbf
  {\bibinfo {volume} {121}},\ \bibinfo {pages} {254501} (\bibinfo {year}
  {2018})}\BibitemShut {NoStop}%
\end{thebibliography}
%

%

\end{document}